\documentclass{llncs}

\usepackage{mathptmx}
\usepackage{helvet}
\usepackage{courier}
\usepackage{multicol}
\usepackage[bottom]{footmisc}
\usepackage{amssymb}
\usepackage{rotating}

\usepackage{epsfig}
\usepackage{latexsym}
\usepackage{graphicx}
\usepackage{color}
\usepackage{url}
\usepackage{amsfonts}
\usepackage{pgf,pgfarrows,pgfnodes,pgfautomata,pgfheaps,pgfshade}
\usepackage{tikz}

\usetikzlibrary{arrows,decorations.pathmorphing,backgrounds,positioning,fit,petri}
\usepackage{etex}

\usepackage{stmaryrd}
\usepackage{amsmath}

\usepackage{bussproofs}

\newcommand{\post}[1]{\mbox{$#1^{\bullet}$}}
\newcommand{\pre}[1]{\mbox{$^{\bullet}#1$}}

\newcommand{\cons}{\mathcal{C}}

\newcommand{\fine}{{\mbox{ }\nolinebreak\hfill{$\Box$}}}

\newcommand{\dec}{\mbox{$dec$}}

\newcommand{\deriv}[1]{{\mbox{${\:\stackrel{#1}{\longrightarrow}\:}$}}}

\newcommand{\Deriv}[1]{{\mbox{${\:\stackrel{#1}{\Longrightarrow}\:}$}}}

\newcommand{\NDeriv}[1]{{\mbox{${\:\stackrel{#1}{\nRightarrow}\:}$}}}
\newcommand{\nderiv}[1]{\nrightarrow}

\newcommand{\eqdef}{ \doteq }

\newcommand{\bigfrac}[2]{
\renewcommand{\arraystretch}{1.5}
\begin{array}{c}#1\\
\hline
#2
\end{array}}

\newcommand{\para}{\mbox{$\,|\,$}}

\newcommand{\nil}{\mbox{\bf 0}}

\newcommand{\const}[1]{\mbox{{\it Const}$(#1)$}}

\renewcommand{\mid}{\;\big|\;}

\newcommand{\encodings}[2]{\ensuremath{\llbracket #2 \rrbracket_{#1}}}

\newcommand{\nat}{{\mathbb N}}

\begin{document}

 \pagestyle{headings}

\title{Distributed Non-Interference
}
\author{Roberto Gorrieri\\
\institute{Universit\`a di Bologna, Dipartimento di Informatica --- Scienza e Ingegneria\\
Mura A. Zamboni, 7,
40127 Bologna, Italy}
\email{{\small roberto.gorrieri@unibo.it}}
}

\maketitle

\begin{abstract}
Information flow security properties were defined some years ago 
(see, e.g., the surveys \cite{FG01,Ry01}) 
in terms of suitable equivalence checking problems. These definitions were provided by using sequential models of 
computations (e.g.,
labeled transition systems \cite{GV15}), and interleaving behavioral equivalences (e.g., bisimulation equivalence \cite{Mil89}). 
More recently, the distributed model of Petri nets has been used to study non-interference in \cite{BG03,BG09,BC15}, but also in these 
papers an interleaving semantics
was used. We argue that in order to capture all the relevant information flows, truly-concurrent behavioral equivalences 
must be used. In particular, 
we propose for Petri nets the distributed non-interference property, called DNI, based on 
{\em branching place bisimilarity} \cite{Gor23b,Gor21b}, which is a sensible, decidable equivalence for finite Petri nets with silent moves. 
Then we focus our attention on the subclass of Petri nets called {\em finite-state machines}, which can be represented (up to isomorphism) by 
the simple process algebra CFM \cite{Gor17}.
DNI is very easily checkable on CFM processes, as it is compositional, so that it does does not 
suffer from the state-space explosion problem. Moreover, we show that DNI can be characterized syntactically on 
CFM by means of a type system.
\end{abstract}

%
\section{Introduction}
%

Non-interference has been defined in the literature as an
extensional property based on some observational semantics: the high
part (usually considered the secret part) of a system does not interfere 
with the low part (i.e., the public one) if whatever is
done at the high level (i.e., by high-level users) produces {\em no visible effect} on the low
part of the system (i.e., it does not change the observations made by low-level users); or, 
equivalently, if low-users cannot infer the occurrence of 
those events that should be observable only by high-users.

The original notion of non-interference in \cite{GM} was defined, using {\em  trace semantics}, for deterministic
automata with outputs. Generalized notions of non-interference were
then designed for more general sequential models of computation, such as (nondeterministic) labeled transition
systems (LTSs, for short), by exploiting also finer notions of observational semantics such as
{\em bisimulation equivalence} (see, e.g., \cite{Ry01,FG95,RS,FG01}). 
The security
properties studied in these papers are all based on the dynamics of systems, as they
are defined by means of one (or more) equivalence check(s). Therefore,
non-interference checking is as difficult as {\em equivalence checking}, a
well-studied hard problem in automata theory and concurrency theory (see, e.g., \cite{HMU01,GV15}).

More recently, the distributed model of Petri nets \cite{Pet81,Rei85,Gor17} was used to study non-interference 
in, e.g., \cite{BG03,BG09,BC15}, but also in these papers the security properties of interest are based 
on {\em interleaving} (i.e., sequential) behavioral semantics  
and the distributed model is used only to show that, under some conditions, these (interleaving) information flows 
can be characterized by the presence (or absence) of certain causal or conflict structures in the net.

The thesis of this paper is that, for security analysis,
it is necessary to describe the behavior of distributed systems by means of a distributed model of computation, 
such as a Petri net, but also to {\em observe} the distributed model by means of some {\em truly-concurrent} behavioral semantics, i.e., 
a semantics that can observe the parallelism of system components or, better, the causality among the actions performed.
There is a wide range of possible truly-concurrent equivalences (see, e.g., \cite{vGG89,vG15} for a 
partial overview) and it may be not obvious to understand which is more suitable. Our intuition is 
that, in order to capture all the possible information flows, it is necessary that the observational semantics
is very concrete, observing not only the partial order of events that have occurred, as in 
{\em fully-concurrent bisimilarity} \cite{BDKP91}, but also 
the structure of the distributed state, as in {\em place bisimilarity} \cite{ABS91}, a sensible behavioral equivalence, which was recently
proved decidable \cite{Gor21} for finite Petri nets.

The non-interference problem can be summarized as follows.
Our aim is to analyze systems that can perform two kinds of actions: {\em high-level} actions, representing the 
interaction of the system with high-level users, and {\em low-level} actions, representing the interaction with 
low-level users. We want to verify whether the interplay between the high user and the high part of the system 
can affect the view of the system as observed by a low user. 
We assume that the low user knows the structure of the system, and we check whether, in spite of this, (s)he is not able 
to infer the behavior of the high user by observing the low view of the execution of the system. 
Hence, we assume that the set of actions is partitioned into two subsets: the set $H$ of high-level actions (or secret actions)
and the set $L$ of low-level actions (or public actions).

To explain our point of view, we use the process algebra CFM \cite{Gor17,Gor19b}, 
extending finite-state CCS \cite{Mil89} with an operator of asynchronous (i.e., without 
communication capabilities) parallelism, that can be used at the top level only. 
The net semantics of CFM \cite{Gor17,Gor19b}, described in Section \ref{net-sem-sec}, 
 associates a {\em finite-state machine} (a particular Petri net whose transitions 
have singleton pre-set and singleton, or empty, post-set) $N_p$ to each CFM process term $p$ and, conversely, for each finite-state machine 
$N$ we can associate a CFM process term $p$ whose net semantics $N_p$ is isomorphic to $N$; for this reason, 
we can work equivalently on CFM process terms or on finite-state machines.
Consider the CFM sequential process

\begin{figure}[t]
	\centering

\begin{tikzpicture}
\def\eodist{0.7cm}
\def\eodisty{2.4cm}
\def\eolab{0.5mm}
\def\eodiaglabeldist{0.5mm}

  \tikzset{%
    mythick/.style={%
        line width=.35mm,>=stealth
    }
  }

  \tikzset{%
    mynode/.style={
      circle,
      fill,
      inner sep=2.1pt
    },
    shorten >= 3pt,
    shorten <= 3pt
  }


\node (a) [label=left:$a)\qquad $]{};

  \node (A) [mynode,label={[label distance=\eodiaglabeldist]left:$A$}] {};
  \node (B) [mynode, below=\eodist of A, label={[label distance=\eodiaglabeldist]left:$B$}] {};

  \draw (A) edge[mythick,->] node[left=\eolab] {$h$} (B);
 \draw (A) edge[loop right,mythick,->] node[right=\eolab] {$l$} (A);
 \draw (B) edge[loop right,mythick,->] node[right=\eolab] {$l$} (B);


\node (b) [right={1.7cm} of a, label=left:$b)\;\;$] {};

  \node (A1) [mynode,right=\eodisty of A, label={[label distance=\eodiaglabeldist]left:$A\setminus h$}] {};
  \node (B1) [mynode, below=\eodist of A1, label={[label distance=\eodiaglabeldist]left:$B \setminus h$}] {};

 \draw (A1) edge[loop right,mythick,->] node[right=\eolab] {$l$} (A1);
 \draw (B1) edge[loop right,mythick,->] node[right=\eolab] {$l$} (B1);


\node (c) [right={2.3cm} of b, label=left:$c)\;\;$] {};

 \node (A2) [mynode,right=\eodisty of A1, label={[label distance=\eodiaglabeldist]left:$C \para B$}] {};
  \node (B2) [mynode, below=\eodist of A2, label={[label distance=\eodiaglabeldist]left:$B \para B$}] {};

  \draw (A2) edge[mythick,->] node[left=\eolab] {$h$} (B2);
 \draw (A2) edge[loop right,mythick,->] node[right=\eolab] {$l$} (A2);
 \draw (B2) edge[loop right,mythick,->] node[right=\eolab] {$l$} (B2);

\end{tikzpicture}

	\caption{An example of a secure process with never-ending behavior}
	\label{prima-fig}
\end{figure}
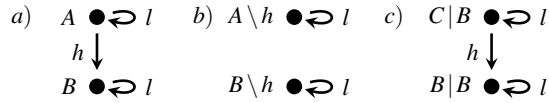

 $\begin{array}{rcllrcl}
 A & \eqdef & l.A + h.B & \qquad& B & \eqdef & l.B
 \end{array}
 $
 
 \noindent
 where $A$ and $B$ are process constants, each one
 equipped with a defining equation, and where $l$ is a low action and $h$ is a 
 high one. The LTS semantics for $A$ is outlined in Figure \ref{prima-fig}(a).

   Intuitively, this system is secure because the execution of the high-level action $h$ does not add any 
 information to a low-level observer: what such a low observer can see is just a, possibly never-ending, 
 sequence of low actions $l$ in any case: {\em before} $h$ is performed as well as {\em after} $h$ has been performed. 
 The most restrictive non-interference property discussed in \cite{FG95,FG01,BG09}, called SBNDC 
 (acronym of {\em Strong Bisimulation Non-Deducibility on Composition}), requires that
whenever the system under scrutiny 
performs a high-level action, the states of the system {\em before} and {\em after} 
executing that high-level action are indistinguishable for a low-level observer. 
 In our example, this means that $A$ is SBNDC if $A \setminus h$ is bisimilar to $B\setminus h$,
 where $A \setminus h$ denotes the low observable part of the system before $h$ is performed (i.e., 
 process $A$ where the transition $h$ is pruned), while $B \setminus h$ denotes the low observable part of the system after $h$ is performed.
 By observing the LTSs in Figure \ref{prima-fig}(b),
 we conclude that $A$ is SBNDC.

 However, the LTS in Figure \ref{prima-fig}(a) is isomorphic to the LTS in \ref{prima-fig}(c), which is the semantics 
 of the parallel process $C \para B$, where
 $C  \eqdef  h.B$ and $B  \eqdef l.B$. Therefore, we can conclude that also $C \para B$ enjoys the SBNDC non-interference
 property. Unfortunately, this parallel
 system is not secure: if a low observer can realize that two occurrences of actions $l$ have been 
 performed in parallel (which is a possible behavior of $B\para B$ only, but this behavior is not represented in the interleaving 
 LTS semantics),
 then (s)he is sure that the high-level action $h$ has been performed. This trivial example shows that 
 SBNDC, based on interleaving bisimulation equivalence, is unable to detect a real information flow 
 from high to low.
 
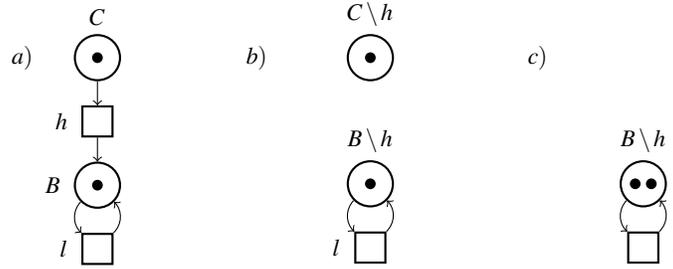
\begin{figure}[t]
	\centering

\begin{tikzpicture}[
every place/.style={draw,thick,inner sep=0pt,minimum size=6mm},
every transition/.style={draw,thick,inner sep=0pt,minimum size=4mm},
bend angle=45,
pre/.style={<-,shorten <=1pt,>=stealth,semithick},
post/.style={->,shorten >=1pt,>=stealth,semithick}
]
\def\eofigdist{3cm}
\def\eodist{0.32}
\def\eodisty{0.7}
\def\eodistw{1.05}

\node (a) [label=left:$a)\qquad $]{};

\node (q1) [place,tokens=1] [label={above:$C$} ] {};
\node (t1) [transition] [below=\eodist of q1,label=left:$h\;$] {};
\node (q2) [place,tokens=1] [below=\eodist of t1,label=left:$B\;$] {};
\node (t2) [transition] [below =\eodist of q2,label=left:$l\;$] {};

\draw  [->] (q1) to (t1);
\draw  [->] (t1) to (q2);
\draw  [->, bend right] (q2) to (t2);
\draw  [->, bend right] (t2) to (q2);


\node (b) [right={2.4cm} of a, label=left:$b)\;\;$] {};

\node (p1) [place,tokens=1]  [right=\eofigdist of q1,label=above:$C \setminus h$] {};
\node (p2) [place,tokens=1] [below=\eodistw of p1,label=above:$B \setminus h$] {};
\node (s2) [transition] [below =\eodist of p2,label=left:$l\;$] {};

\draw  [->, bend right] (p2) to (s2);
\draw  [->, bend right] (s2) to (p2);


\node (c) [right={3.5cm} of b,label=left:$c)\;\;$] {};

\node (r1) [place, tokens=2]  [right=\eofigdist of p2,label=above:$B \setminus h$] {};
\node (v1)  [transition] [below=\eodist of r1,label=right:$\;l$] {};

\draw  [->, bend right] (r1) to (v1);
\draw  [->, bend right] (v1) to (r1);

\end{tikzpicture}

	\caption{The Petri net for the process $C \para B$ in (a), of $(C \para B) \setminus h$ in (b) and
	of $(B \para B) \setminus h$ in (c)}
	\label{seconda-fig}
\end{figure}

 However, if we use a Petri net semantics for the process algebra CFM, as described in Section \ref{net-sem-sec},
  we have that the net semantics for $C \para B$, outlined in Figure \ref{seconda-fig}(a), is such that
 the low-observable system {\em before} the execution of $h$, i.e., $(C \para B)\setminus h$ (whose net semantics is outlined in 
 Figure \ref{seconda-fig}(b)) is not ``truly-concurrent'' equivalent 
 to the low-observable system {\em after} $h$, i.e., $(B \para B) \setminus h$ (whose net semantics is in Figure \ref{seconda-fig}(c)). In fact,
 it is now visible that the two tokens on place $B \setminus h$ in the net in (c) can perform transition $l$ at the same time, 
 and such a behavior is impossible for the net in (b). 
 This  example calls for a truly-concurrent semantics which is 
 at least as discriminating as {\em step
bisimulation} equivalence \cite{NT84,Gor17}, which is able to observe the parallel execution of actions.

\begin{figure}[t]
	\centering

\begin{tikzpicture}[
every place/.style={draw,thick,inner sep=0pt,minimum size=6mm},
every transition/.style={draw,thick,inner sep=0pt,minimum size=4mm},
bend angle=45,
pre/.style={<-,shorten <=1pt,>=stealth,semithick},
post/.style={->,shorten >=1pt,>=stealth,semithick}
]
\def\eofigdist{3.8cm}
\def\eofigdisty{2.8cm}
\def\eodist{0.4}
\def\eodisty{0.5}
\def\eodistw{1}
\def\eodistz{2}

\node (a) [label=left:$a)\qquad \quad $]{};

\node (q1) [place,tokens=1] [label={above:$A$} ] {};
\node (t1) [transition] [below =\eodistw of q1,label=left:$h$] {};
\node (t2) [transition] [below right=\eodistw of q1,label=left:$a$] {};
\node (t3) [transition] [below left=\eodistw of q1,label=left:$b$] {};
\node (q2) [place] [below left=\eodisty of t1,label=left:$a.\nil$] {};
\node (q3) [place] [below right=\eodisty of t1,label=left:$b.\nil$] {};
\node (t4) [transition] [below =\eodist of q2,label=left:$a$] {};
\node (t5) [transition] [below =\eodist of q3,label=right:$b$] {};

\draw  [->] (q1) to (t1);
\draw  [->] (q1) to (t2);
\draw  [->] (q1) to (t3);
\draw  [->] (t1) to (q2);
\draw  [->] (t1) to (q3);
\draw  [->] (t2) to (q3);
\draw  [->] (t3) to (q2);
\draw  [->] (q2) to (t4);
\draw  [->] (q3) to (t5);


\node (b) [right={3.2cm} of a, label=left:$b)\;\;$] {};

\node (p1) [place,tokens=1]  [right=\eofigdist of q1,label=above:$A \setminus h$] {};
\node (s2) [transition] [below right=\eodistw of p1,label=left:$a$] {};
\node (s3) [transition] [below left=\eodistw of p1,label=right:$b$] {};
\node (p4) [place] [below=\eodist of s2,label=right:$b.\nil \setminus h$] {};
\node (p5) [place] [below=\eodist of s3,label=right:$a.\nil \setminus h$] {};
\node (s4) [transition] [below =\eodist of p4,label=right:$b$] {};
\node (s5) [transition] [below =\eodist of p5,label=right:$a$] {};

\draw  [->] (p1) to (s2);
\draw  [->] (p1) to (s3);
\draw  [->] (s2) to (p4);
\draw  [->] (s3) to (p5);
\draw  [->] (p4) to (s4);
\draw  [->] (p5) to (s5);


\node (c) [right={3.6cm} of b,label=left:$c)\;\;$] {};

\node (r2) [place, tokens=1]  [right=\eofigdisty of p1,label=above:$a.\nil \setminus h$] {};
\node (r3) [place, tokens=1]  [right=\eodist of r2,label=above:$b.\nil \setminus h$] {};
\node (v3)  [transition] [below=\eodist of r2,label=right:$a$] {};
\node (v4)  [transition] [below=\eodist of r3,label=right:$b$] {};

\draw  [->] (r2) to (v3);
\draw  [->] (r3) to (v4);

\end{tikzpicture}

	\caption{An example of an insecure process}
	\label{terza-fig}
\end{figure}
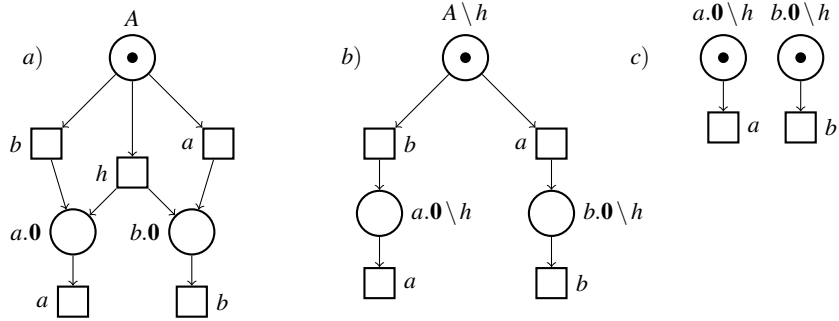

 \begin{figure}[t]
	\centering

\begin{tikzpicture}[
every place/.style={draw,thick,inner sep=0pt,minimum size=6mm},
every transition/.style={draw,thick,inner sep=0pt,minimum size=4mm},
bend angle=45,
pre/.style={<-,shorten <=1pt,>=stealth,semithick},
post/.style={->,shorten >=1pt,>=stealth,semithick}
]
\def\eofigdist{3cm}
\def\eodist{0.32}
\def\eodisty{0.52}
\def\eodistw{1.1}

\node (a) [label=left:$a)\qquad \quad $]{};

\node (q1) [place,tokens=1] [label={above:$l.h.C$} ] {};
\node (t1) [transition] [below=\eodist of q1,label=left:$l\;$] {};
\node (q2) [place] [below=\eodist of t1,label=left:$h.C$] {};
\node (t2) [transition] [below =\eodist of q2,label=left:$h\;$] {};
\node (q3) [place] [below =\eodist of t2,label=left:$C$] {};

\draw  [->] (q1) to (t1);
\draw  [->] (t1) to (q2);
\draw  [->] (q2) to (t2);
\draw  [->] (t2) to (q3);


\node (b) [right={2.8cm} of a, label=left:$b)\;\;$] {};

\node (p1) [place,tokens=1]  [right=\eofigdist of q1,label=above:$l.h.\nil$] {};
\node (s2) [transition] [below =\eodist of p1,label=right:$l$] {};
\node (p3) [place] [below =\eodist of s2,label=right:$h.\nil$] {};
\node (s4) [transition] [below =\eodist of p3,label=right:$h$] {};

\draw  [->] (p1) to (s2);
\draw  [->] (s2) to (p3);
\draw  [->] (p3) to (s4);


\node (c) [right={3.3cm} of b,label=left:$c)\;\;$] {};

\node (r1) [place, tokens=1]  [right=\eofigdist of p1,label=above:$h.l.\nil + l.C$] {};
\node (v1)  [transition] [below left=\eodisty of r1,label=right:$h$] {};
\node (v2)  [transition] [below right=\eodisty of r1,label=right:$l$] {};
\node (r7) [place] [below =\eodist of v1,label=right:$l.\nil$] {};
\node (r8) [place] [below =\eodist of v2,label=right:$C$] {};
\node (v3)  [transition] [below=\eodist of r7,label=right:$l$] {};

\draw  [->] (r1) to (v1);
\draw  [->] (r1) to (v2);
\draw  [->] (v1) to (r7);
\draw  [->] (v2) to (r8);
\draw  [->] (r7) to (v3);

\end{tikzpicture}

	\caption{By observing the state, further information flows can be detectable}
	\label{quarta-fig}
\end{figure}
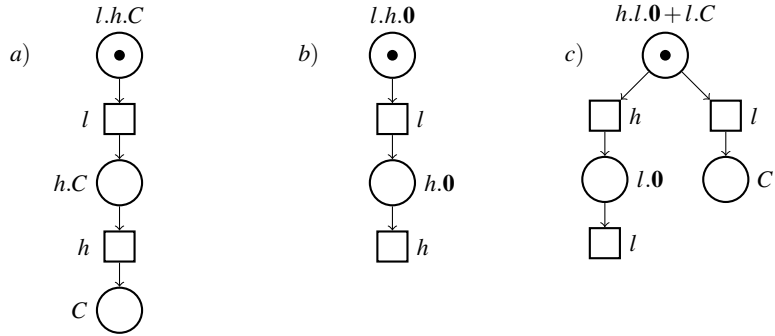

As a further example, consider the process term (not expressible in CFM, rather in the slightly richer process algebra
BPP \cite{Ch93,GV15,Gor22-tcs}):

 $\begin{array}{rcllrcl}
 A & \eqdef & h.(a.\nil | b.\nil) + a.b.\nil + b.a.\nil
 \end{array}
 $

\noindent
where $a, b$ are low-level actions, whose net semantics is outlined in Figure \ref{terza-fig}(a). 
The two low-observable subnets {\em before} and {\em after} the execution 
of $h$ (depicted in Figure \ref{terza-fig}(b) and \ref{terza-fig}(c), respectively) generate different partial orders: 
if an observer realizes that $a$ and $b$ are causally 
independent, then (s)he infers that
$h$ has been performed. 
Therefore, the observation of {\em causality} is crucial for detecting potential information flows in a distributed system; hence, 
our observational semantics
should be at least as discriminating as {\em fully-concurrent bisimilarity} \cite{BDKP91} (or, equivalently, {\em history-preserving bisimilarity} \cite{RT88,vGG89,DDM89}), that is able to observe the partial orders of performed events.

However, we want to argue that also the structure of the distributed system (i.e, the size of the current marking, 
in Petri net terminology) 
 is an important feature, that cannot be ignored, as most truly-concurrent behavioral equivalences do 
 (notably fully-concurrent bisimilarity \cite{BDKP91}).  
Consider the net in Figure \ref{quarta-fig}(a), which is
the semantics of the CFM process $l.h.C$, with $C \eqdef \nil$. 
Since $h$ is not causing any low action, we conclude that
this net is secure. However, the very similar net in (b), which is the semantics of $l.h.\nil$, is not 
secure: if the low observer realizes that the token
disappears in the end, then (s)he can infer that $h$ has been performed. This simple observation is about the
observability of the size of the distributed system: if the execution of a high-level action modifies the current 
number of tokens (i.e., the number of currently active sequential subprocesses),
then its execution has an observable effect that can be recognized by a low observer.
Indeed, also in the net in Figure \ref{terza-fig} the execution of the high-level transitions $h$ modifies the current number of active components,
because the marking before executing $h$ has size $1$, while the marking after $h$ has size $2$, so that a low observer can realize that
$h$ has been performed.
Similarly, the net in Figure \ref{quarta-fig}(c), which is the semantics of $h.l.\nil + l.C$ with $C \eqdef \nil$, is such that, before and after $h$, 
the same partial order of events can be performed, but
if the token disappears in the end, then the low observer knows that $h$ has been performed.
Therefore, it is necessary to use a truly-concurrent behavioral equivalence slightly finer than 
fully-concurrent bisimilarity, which is {\em resource-sensitive}, i.e., able to observe also the resources of the distributed state
(i.e., the number of sequential processes composing the distributed system).
Resource-sensitive truly-concurrent behavioral equivalences include, e.g., 
{\em place bisimilarity} \cite{ABS91,Gor21} and
{\em structure-preserving bisimilarity} \cite{vG15}. These equivalences coincide on CFM (i.e., on finite-state machines) and, on this class of nets,
they can be equivalently characterized in a very simple and effective way as {\em team bisimilarity} \cite{Gor17b,Gor22-tcs}.
However, on finite Petri nets place bisimilarity is strictly finer than structure-preserving bisimilarity, 
but the former is decidable \cite{Gor21},
while this is an open problem for the latter. (By the way, fully-concurrent bisimilarity is undecidable for finite Petri nets \cite{Esp98}.)
So, our starting point is to consider {\em place bisimilarity}
 as our candidate observational semantics over finite Petri nets.

When considering Petri nets with {\em silent} transitions (i.e., transitions labeled by the invisible action $\tau$), 
we have the problem to understand which kind of generalization of
place bisimulation equivalence to consider. We have at least two possible alternatives, inspired to {\em weak bisimilarity} \cite{Mil89}
or to {\em branching bisimilarity} \cite{vGW96}, originally proposed for LTSs \cite{GV15}. 
In order to understand the problem, consider the CFM
process constants

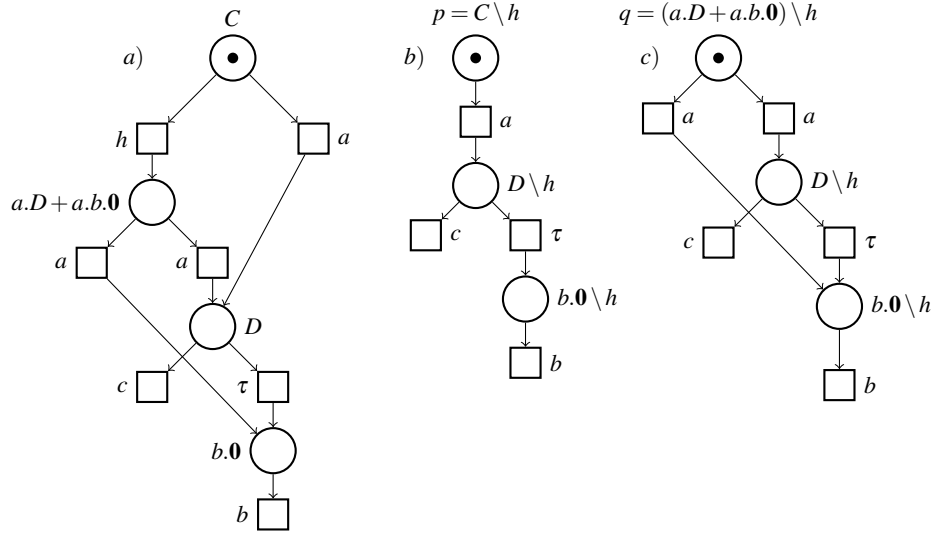
\begin{figure}[t]
	\centering

\begin{tikzpicture}[
every place/.style={draw,thick,inner sep=0pt,minimum size=6mm},
every transition/.style={draw,thick,inner sep=0pt,minimum size=4mm},
bend angle=45,
pre/.style={<-,shorten <=1pt,>=stealth,semithick},
post/.style={->,shorten >=1pt,>=stealth,semithick}
]
\def\eofigdist{2.6cm}
\def\eofigdisty{3cm}
\def\eodist{0.32}
\def\eodisty{0.52}
\def\eodistw{0.9}

\node (a) [label=left:$a)\qquad \quad $]{};

\node (q1) [place,tokens=1] [label={above:$C$} ] {};
\node (t1) [transition] [below left=\eodistw of q1,label=left:$h$] {};
\node (t2) [transition] [below right=\eodistw of q1,label=right:$a$] {};
\node (q2) [place] [below=\eodist of t1,label=left:$a.D+a.b.\nil$] {};
\node (t3) [transition] [below left =\eodisty of q2,label=left:$a$] {};
\node (t4) [transition] [below right =\eodisty of q2,label=left:$a$] {};
\node (q3) [place] [below =\eodist of t4,label=right:$D$] {};
\node (t5) [transition] [below left =\eodisty of q3,label=left:$c$] {};
\node (t6) [transition] [below right =\eodisty of q3,label=left:$\tau$] {};
\node (q4) [place] [below =\eodist of t6,label=left:$b.\nil$] {};
\node (t7) [transition] [below =\eodist of q4,label=left:$b$] {};

\draw  [->] (q1) to (t1);
\draw  [->] (q1) to (t2);
\draw  [->] (t1) to (q2);
\draw  [->] (t2) to (q3);
\draw  [->] (q2) to (t3);
\draw  [->] (q2) to (t4);
\draw  [->] (t4) to (q3);
\draw  [->] (q3) to (t5);
\draw  [->] (q3) to (t6);
\draw  [->] (t3) to (q4);
\draw  [->] (t6) to (q4);
\draw  [->] (q4) to (t7);


\node (b) [right={2.7cm} of a, label=left:$b)\;\;$] {};

\node (p1) [place,tokens=1]  [right=\eofigdist of q1,label={above:$p = C\setminus h$}] {};
\node (s2) [transition] [below =\eodist of p1,label=right:$a$] {};
\node (p3) [place] [below =\eodist of s2,label=right:$D\setminus h$] {};
\node (s4) [transition] [below left =\eodist of p3,label=right:$c$] {};
\node (s5) [transition] [below right =\eodist of p3,label=right:$\tau$] {};
\node (p4) [place] [below =\eodist of s5,label=right:$b.\nil \setminus h$] {};
\node (s6) [transition] [below =\eodist of p4,label=right:$b$] {};

\draw  [->] (p1) to (s2);
\draw  [->] (s2) to (p3);
\draw  [->] (p3) to (s4);
\draw  [->] (p3) to (s5);
\draw  [->] (s5) to (p4);
\draw  [->] (p4) to (s6);


\node (c) [right={2.9cm} of b,label=left:$c)\;\;$] {};

\node (r1) [place, tokens=1]  [right=\eofigdist of p1,label={above:$q = (a.D +a.b.\nil)\setminus h$}] {};
\node (v1)  [transition] [below left=\eodisty of r1,label=right:$a$] {};
\node (v2)  [transition] [below right=\eodisty of r1,label=right:$a$] {};
\node (r2) [place] [below =\eodist of v2,label=right:$D \setminus h$] {};
\node (v3)  [transition] [below left=\eodisty of r2,label=left:$c$] {};
\node (v4)  [transition] [below right=\eodisty of r2,label=right:$\tau$] {};
\node (r3) [place] [below =\eodist of v4,label=right:$b.\nil \setminus h$] {};
\node (v5)  [transition] [below =\eodisty of r3,label=right:$b$] {};

\draw  [->] (r1) to (v1);
\draw  [->] (r1) to (v2);
\draw  [->] (v2) to (r2);
\draw  [->] (r2) to (v3);
\draw  [->] (r2) to (v4);
\draw  [->] (v4) to (r3);
\draw  [->] (r3) to (v5);
\draw  [->] (v1) to (r3);

\end{tikzpicture}

	\caption{By observing the timing of choice, further information flows can be detected}
	\label{quinta-fig}
\end{figure}

$\begin{array}{rcllrcl}
 C & \eqdef & h.(a.D + a.b.\nil) + a.D & \qquad & D & \eqdef & \tau.b.\nil + c.\nil
 \end{array}
 $

\noindent
where $a, b, c$ are low-level actions and $h$ is a high-level action, whose net semantics is outlined in 
Figure \ref{quinta-fig}(a). Of course, the low-observable system before executing $h$ is 
$p = C \setminus h$, whose net is outlined in Figure \ref{quinta-fig}(b), while the low-observable system after 
$h$ is $q= (a.D + a.b.\nil)\setminus h$,
depicted in Figure \ref{quinta-fig}(c).

According to weak
bisimilarity, $p$ and $q$ are equivalent; in particular, to the transition $q \deriv{a} b.\nil \setminus h$, process 
$p$ can reply with the sequence
$p \deriv{a} D \setminus h \deriv{\tau} b.\nil \setminus h$.
Therefore, if we use weak bisimilarity, then the system would be considered secure. However,
note that, by replying to the transition $q \deriv{a} b.\nil \setminus h$, the process $p$ passes through state $D \setminus h$,
that is capable of performing $c$, a behavior that is impossible for $b.\nil \setminus h$.
Hence, if a low user realizes that,
after performing the low action $a$, the low action $c$ is never available (i.e., it is able to realize that the user is
executing the computation due to summand $a.b.\nil$ as in 
$q \deriv{a} b.\nil \setminus h$),
then (s)he is sure that $h$ has been performed; hence, $C$ cannot be considered as secure. 

As a matter of fact, 
branching bisimilarity, which is strictly finer than weak bisimilarity, 
does not consider $p$ and $q$ as equivalent, as only $q$ has the possibility of discarding $c$ immediately 
by performing $a$. 
In fact, this crucial property is enjoyed by branching bisimilarity (but not by weak bisimilarity):
when in the branching bisimulation game a transition $q_1 \deriv{\mu} q_1'$ is matched by a computation, 
say, $q_2 \Deriv{\epsilon} q_2' \deriv{\mu} q_2'' \Deriv{\epsilon} q_2'''$, 
all the states traversed by the silent computation from $q_2$ to $q_2'$ are 
branching bisimilar, so that they all belong to the same equivalence class; and, similarly,
all the states traversed by the silent computation from $q_2''$ to $q_2'''$ are branching bisimilar (in the example above, 
the states $D \setminus h$ and $b.\nil \setminus h$ are not equivalent, as only the former can perform $c$).
This ensures that
branching bisimilarity does fully respect the timing of choices.

Indeed, this example reinforces the observation that the {\em timing of choices} is a crucial aspect for security,
as already observed on LTSs in \cite{FG95,FG01} where it is argued that it is better to use 
branching-time behavioral equivalences (such as {\em bisimulation} equivalence \cite{Mil89})
rather than linear-time behavioral equivalences (such as {\em trace} equivalence).

Summarizing, the behavioral equivalence that we advocate is {\em branching place bisimilarity} \cite{Gor23b,Gor21b}, 
a sensible behavioral equivalence, which
is decidable in exponential time on finite Petri nets and that specializes to the very efficiently decidable {\em branching 
team equivalence} on finite-state machines \cite{Gor19b}.

So, we are now ready to propose the property DNI (acronym of {\em Distributed 
Non-Interference}) as follows. Given a finite Petri net $N$, we denote by $N \setminus H$ the net
obtained from $N$ by pruning all the high-labeled transitions and by denoting each place $s$ of $N$ by $s \setminus H$, as we did in the previous examples.
\begin{quote}
The initial marking $m_0$ of $N$ is {\bf DNI}
if for each reachable markings $m_1, m_2$ and for each $h \in H$ such that $m_1 \deriv{h}m_2$, we have that $m_1 \setminus H \approx_p m_2 \setminus H$, where $\approx_p$ denotes branching
place bisimilarity and, for $i = 1 ,2$,
$m_i \setminus H$ denotes the marking of the net $N \setminus H$ corresponding to $m_i$ in $N$.
\end{quote}

On the one hand, DNI is clearly decidable in exponential time for bounded nets (i.e., nets with a finite number of reachable markings),
because we have to make a finite number of branching place bisimulation checks, each one decidable in exponential time \cite{Gor21b}.
On the other hand, it is an open problem to see
whether DNI is decidable for general unbounded Petri nets. 

After this general introduction, in the rest of the paper, we focus our attention only on the special class of 
Petri nets called {\em finite-state machines}, on the corresponding
process algebra CFM that truly represents such nets, up to isomorphism \cite{Gor17}, and on {\em branching team equivalence} \cite{Gor19b}, 
which is the specialization of branching place bisimilarity on such nets.
Hence, the definition of DNI above can be specialized for this subcase in the obvious way.

On the class of finite-state machines, which are bounded, 
DNI is very easily checkable. As a matter of fact, given a CFM parallel process $p = p_1 \para p_2 \para \ldots \para p_n$, 
we prove that $p$ is DNI if and only if each $p_i$ is DNI, for $i = 1, \ldots, n$. Hence,
instead of inspecting the state space of $p$, we can simply inspect the state spaces for $p_1$, $p_2$, \ldots $p_n$.
If the state space of each $p_i$ is composed of 10 states, then the state space of $p$ is composed of 
$10^n$ states, so that a direct check of DNI on the state space of $p$ 
is impossible for large values of $n$. However,
the distributed analysis on each $p_i$ can be done in linear time w.r.t. $n$, hence also for very large values of $n$, so that the 
state-space explosion problem can be kept under control.

Moreover, a structural characterization of DNI can be provided by inspecting directly the finite-state machine under scrutiny, 
which offers a very efficient, polynomial algorithm
to check DNI. 

Finally, a slight enhancement of DNI (called DNI as well, with abuse of notation), based on the slightly finer 
{\em rooted} branching team equivalence \cite{Gor19b} (which is the coarsest congruence for all the operators of CFM contained in branching team equivalence), is characterized syntactically 
on CFM by means of a typing system: we prove that
a CFM process $p$ is DNI if and only if $p$ (or a slight variant of it) is typed. The typing system is based on
 a finite, sound and complete, axiomatization of rooted branching team equivalence \cite{Gor19b}.

The paper is organized as follows. Section \ref{def-bsec} introduces the basic definitions about finite-state machines, 
together with the definition 
of (rooted) branching team  equivalence.
Section \ref{cfm-sec} defines the syntax of the process algebra CFM, hints its net semantics and recalls the 
finite axiomatization of rooted branching team equivalence from \cite{Gor19b}.
Section \ref{dni-sec} introduces the distributed non-interference property DNI on CFM processes, 
proving that it can be really 
checked in a distributed manner, and also describes the typing system for (the slightly finer variant of) DNI (based on the rooted variant
of branching team bisimulation).
Finally, Section \ref{conc-bsec}  comments on related work.

%
\section{Finite-State Machines and Branching Team Equivalence} \label{def-bsec}
%

By finite-state machine (FSM, for short) 
we mean a simple type of finite Petri net \cite{Pet81,Rei85,Gor17} whose transitions have singleton 
pre-set and singleton, or empty, post-set. 
The name originates from the fact that an unmarked net of this kind is essentially isomorphic to
a nondeterministic finite automaton \cite{HMU01} (NFA, for short), usually called a finite-state machine as well.
However, semantically, our FSMs are richer than NFAs
because, as their initial marking may be not a singleton, these nets can also exhibit
concurrent behavior, while NFAs are strictly sequential.

\begin{definition}\label{multiset}{\bf (Multiset)}\index{Multiset}
Let $\nat$ be the set of natural numbers. 
Given a finite set $S$, a {\em multiset} over $S$ is a function $m: S \rightarrow\nat$. 
Its {\em support} set $dom(m)$ is  $\{ s \in S \mid m(s) \neq 0\}$. 
The set of all multisets 
over $S$ is ${\mathcal M}(S)$, ranged over by $m$. 
We write $s \in m$ if $m(s)>0$.
The {\em multiplicity} of $s$ in $m$ is the number $m(s)$. The {\em size} of $m$, denoted by $|m|$,
is $\sum_{s\in S} m(s)$, i.e., the total number of its elements.
A multiset $m$ such 
that $dom(m) = \emptyset$ is called {\em empty} and is denoted by $\theta$.
We write $m \subseteq m'$ if $m(s) \leq m'(s)$ for all $s \in S$. 
{\em Multiset union} $\_ \oplus \_$ is defined as follows: $(m \oplus m')(s)$ $ = m(s) + m'(s)$.
{\em Multiset difference} $\_ \ominus \_$ is defined as follows: 
$(m_1 \ominus m_2)(s) = max\{m_1(s) - m_2(s), 0\}$.
The {\em scalar product} of a number $j$ with $m$ is the multiset $j \cdot m$ defined as
$(j \cdot m)(s) = j \cdot (m(s))$. By $s_i$ we also denote the multiset with $s_i$ as its only element.
Hence, a multiset $m$ over $S = \{s_1, \ldots, s_n\}$
can be represented as $k_1\cdot s_{1} \oplus k_2 \cdot s_{2} \oplus \ldots \oplus k_n \cdot s_{n}$,
where $k_j = m(s_{j}) \geq 0$ for $j= 1, \ldots, n$.
\fine
\end{definition}

\begin{definition}\label{fsm-net-def}\index{FSM Petri net!definition}{\bf (Finite-state machine)}
A labeled {\em finite-state machine} (FSM, for short) is a tuple $N = (S, A, T)$, where
\begin{itemize}
\item 
$S$ is the finite set of {\em places}, ranged over by $s$ (possibly indexed),
\item 
$A$ is the finite set of {\em labels}, ranged over by $\ell$ (possibly indexed), which contains the silent label $\tau$, and
\item 
$T \subseteq S \times A \times (S \cup \{\theta\})$ 
is the finite set of {\em transitions}, 
ranged over by $t$. 
\end{itemize}

Given a transition $t = (s, \ell, m)$,
we use the notation $\pre t$ to denote its {\em pre-set} $s$ (which is a single place), i.e., one token on place $s$ to be consumed; 
$l(t)$ for its {\em label} $\ell$, and
$\post t$ to denote its {\em post-set} $m$ (which is a place or the empty multiset $\theta$), i.e., the token, if any, to be produced.
Hence, transition $t$ can be also represented as $\pre t \deriv{l(t)} \post t$.
\fine
\end{definition}

Graphically, a place is represented by a  circle, a transition by a  box,
connected by a directed arc from the place in its pre-set and to the place in its post-set, if any.

\begin{definition}\label{net-system}{\bf (Marking, FSM net system)}\index{Marking}\index{Token}\index{FSM net system!definition}\index{FSM net!marked}
A multiset over $S$  is called a {\em marking}. Given a marking $m$ and a place $s$, 
we say that the place $s$ contains $m(s)$ {\em tokens}, graphically represented by $m(s)$ bullets
inside place $s$.
An {\em FSM net system} $N(m_0)$ is a tuple $(S, A, T,$ $ m_{0})$, where $(S,A, T)$ is an FSM and $m_{0}$ is  
a marking over $S$, called
the {\em initial marking}. We also say that $N(m_0)$ is a {\em marked} net. 
An FSM net system $N(m_0) = (S, A, T,$ $ m_0)$ is {\em sequential} if 
$m_0$ is a singleton, i.e., $|m_0| = 1$; while it is {\em concurrent} if
$m_0$ is arbitrary.
\fine
\end{definition}

\begin{definition}\label{firing-system}{\bf (Firing sequence, reachable markings)}
Given an FSM $N = (S, A, T)$, a transition $t $ is {\em enabled} at marking $m$, 
denoted by $m[t\rangle$, if $\pre t \subseteq m$. 
The execution (or {\em firing}) of  $t$ enabled at $m$ produces the marking $m' = (m \ominus  \pre t) \oplus \post t$. 
This is written usually as $m[t\rangle m'$, but also as $m \deriv{l(t)} m'$. 
A {\em firing sequence} starting at $m$ is defined inductively as 
\begin{itemize}
\item $m[\epsilon\rangle m$ is a firing sequence (where $\epsilon$ denotes an empty sequence of transitions) and
\item if $m[\sigma\rangle m'$ is a firing sequence and $m' [t\rangle m''$, then
$m [\sigma t\rangle m''$ is a firing sequence. 
\end{itemize}
If $\sigma = t_1 \ldots t_n$ (for $n \geq 0$) and $m[\sigma\rangle m'$ is a firing sequence, then there exist $m_1,  \ldots, m_{n+1}$ such that
$m = m_1[t_1\rangle m_2 [t_2\rangle \ldots  m_n [t_n\rangle m_{n+1} = m'$, and 
$\sigma = t_1 \ldots t_n$ is called a {\em transition sequence} starting at $m$ and ending 
at $m'$. 
The set of {\em reachable markings} from $m$ is $reach(m) = \{m' \mid \exists \sigma.
m[\sigma\rangle m'\}$.
\fine
\end{definition}

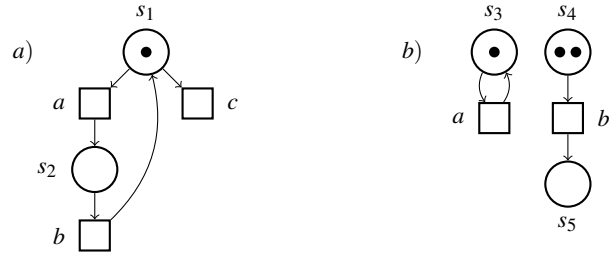
\begin{figure}[t]
\centering

\begin{tikzpicture}[
every place/.style={draw,thick,inner sep=0pt,minimum size=6mm},
every transition/.style={draw,thick,inner sep=0pt,minimum size=4mm},
bend angle=30,
pre/.style={<-,shorten <=1pt,>=stealth,semithick},
post/.style={->,shorten >=1pt,>=stealth,semithick}
]
\def\eofigdist{4cm}
\def\eodist{0.35cm}

\node (p1) [place,tokens=1]  [label=above:$s_1$] {};
\node (t1) [transition] [below left=\eodist of p1,label=left:$a\;$] {};
\node (t2) [transition] [below right=\eodist of p1,label=right:$\;c$] {};
\node (p2) [place] [below=\eodist of t1,label=left:$s_2\;$] {};
\node (t3) [transition] [below=\eodist of p2,label=left:$b\;$] {};

\node (a) [label=left:$a)\qquad \qquad$]{};

\draw  [->] (p1) to (t1);
\draw  [->] (p1) to (t2);
\draw  [->] (t1) to (p2);
\draw  [->] (p2) to (t3);
\draw  [->, bend right] (t3) to (p1);

  
\node (p3) [place,tokens=1]  [right=\eofigdist of p1, label=above:$s_3$] {};
\node (p4) [place,tokens=2]  [right=\eodist of p3,label=above:$s_4$] {};
\node (t4) [transition] [below=\eodist of p3,label=left:$a\;$] {};
\node (t5) [transition] [below=\eodist of p4,label=right:$\;b$] {};
\node (p5) [place]  [below=\eodist of t5,label=below:$s_5$] {};

\draw  [->, bend right] (p3) to (t4);
\draw  [->, bend right] (t4) to (p3);
\draw  [->] (p4) to (t5);
\draw  [->] (t5) to (p5);

\node (b) [right={4cm} of a,label=left:$b)\quad$] {};

\end{tikzpicture}

\caption{A sequential finite-state machine in (a), and a concurrent finite-state machine in (b)}
\label{net-fsm}
\end{figure}

\begin{example}\label{primo-ex}
By using the usual drawing convention for Petri nets, Figure \ref{net-fsm} shows in (a) a sequential FSM, which performs a 
sequence of $a$'s and $b$'s, 
until it performs one $c$ and then stops {\em successfully} (the token disappears in the end). 
A sequential FSM is such that any reachable marking is a singleton or empty.
Hence, a sequential FSM is {\em safe} (or $1$-bounded): each place in any reachable marking can hold one token at most. 
In (b), a concurrent FSM is depicted: it can perform $a$ forever, interleaved with
two occurrences of $b$, only: the two tokens in $s_4$ will eventually reach $s_5$, which is a place
representing unsuccessful termination (deadlock).
A concurrent FSM is {\em $k$-bounded}, where $k$ is the size 
of the initial marking: 
each place in any reachable marking can hold $k$ tokens at most. Hence, the set $reach(m)$
is finite for any $m$.
As a final comment, note that for each FSM $N = (S, A, T)$ and each place $s \in S$,
the set $reach(s)$ is a subset of $S \cup \{\theta\}$.
\fine
\end{example}

%
\subsection{Branching Bisimulation on Places} \label{bhbis-sec}
%

In order to adapt the definition of branching bisimulation on LTSs \cite{vGW96}
for unmarked FSMs, we need some auxiliary notation.
We define relation $\Deriv{\epsilon} \subseteq S \times (S \cup \{\theta\})$ as the reflexive and transitive closure of 
the silent transition relation; formally, $\forall s \in S$,  
$s \Deriv{\epsilon} s$, denoting that each place can silently 
reach itself with zero steps; moreover, if $s \Deriv{\epsilon} s'$ and $s' \deriv{\tau} m$, then $s \Deriv{\epsilon} m$,
where $m$ can be either the empty marking $\theta$ or a single place because of the shape of FSMs transitions.

\begin{definition}\label{def-br-bis}{\bf (Branching bisimulation on places)}
Let $N = (S, A, T)$ be an FSM.
A {\em branching bisimulation} is a relation
$R\subseteq S\times S$ such that if $(s_1, s_2) \in R$
then for all $\ell \in A$
\begin{itemize}
\item $\forall m_1$ such that  $s_1\deriv{\ell} m_1$,
\begin{itemize}
    \item[--]  either $\ell = \tau$ and $\exists m_2$ such that
    $s_2\Deriv{\epsilon} m_2$ with $(s_1, m_2) \in R$ and $(m_1, m_2) \in R$,
     \item[--] or $\exists s, m_2$ such that 
     $s_2 \Deriv{\epsilon} s \deriv{\ell} m_2$ with $(s_1, s) \in R$ and either 
     $m_1 = \theta = m_2$ or $(m_1, m_2) \in R$,
\end{itemize}
\item and, symmetrically, $\forall m_2$ such that  $s_2 \deriv{\ell} m_2$.
\end{itemize}

Two places $s$ and $s'$ are {\em branching bisimilar} (or {\em branching bisimulation equivalent}), 
denoted by $s \approx s'$, if there exists a 
branching bisimulation $R$ such that $(s, s') \in R$.
\fine
\end{definition}

It is an easy exercise to check \cite{Gor19b} that $(i)$ the identity relation ${\mathcal I} = \{ (s, s) \mid s \in S \}$ is a branching bisimulation;
$(ii)$ the inverse relation $R^{-1} = \{ (s', s) \mid (s, s') \in R\}$ of a branching bisimulation $R$ is a branching bisimulation;
$(iii)$ the relational composition $R_1 \circ R_2 = \{ (s, s'') \mid \exists s'. (s, s') \in R_1 \wedge (s', s'') \in R_2 \}$ of
two branching bisimulations $R_1$ and $R_2$ is a branching bisimulation;
$(iv)$ the union $\bigcup_{i \in I} R_i$ of branching bisimulations $R_i$ is a branching bisimulation.
From these observations, it follows that $\approx$ is an equivalence relation.

Remember that  $s \approx s'$ if there exists a branching bisimulation containing the pair $(s, s')$. 
This means that $\approx$ is the union of 
all branching bisimulations, i.e., 
\[\approx \; =  \bigcup \{ R  \subseteq S \times S  \mid R \; \mbox{ is a branching bisimulation} \}.\] 
Hence, as the union of branching bisimulations is a branching bisimulation, $\approx$
is also a branching bisimulation, hence the largest such relation.

\begin{remark}{\bf (Stuttering Property)} \label{stutt-prop-rem}
It is not difficult to prove that, given a $\tau$-labeled path 
$s_1\deriv{\tau} s_2 \deriv{\tau}  \ldots s_n\deriv{\tau} s_{n+1}$, 
if $s_1 \approx s_{n+1}$,
then $s_i \approx s_j$ for all $i, j = 1, \ldots n+1$. This is sometimes called the {\em stuttering property} \cite{vGW96,GV15}.

This property justifies the following observation on the nature of branching bisimilarity. 
As $\approx$ is a branching bisimulation, it satisfies
the conditions in Definition \ref{def-br-bis}.
Let us consider two branching bisimilar places $s_1 \approx s_2$. Then, suppose $s_1 \deriv{\tau} m_1$ and that $s_2$ responds by performing
$s_2\Deriv{\epsilon}m_2$ with $s_1 \approx m_2$ and $m_1 \approx m_2$. By transitivity of $\approx$, we have that also $s_2 \approx m_2$.
Hence, by the stuttering property, $s_1$ is branching bisimilar to each place in the path from $s_2$ to $m_2$,
and so all the places traversed in the path from $s_2$ to $m_2$ are branching bisimilar.
Similarly, assume $s_1 \deriv{\ell} m_1$ (with $\ell$ that can be $\tau$)
and that $s_2$ responds by performing $s_2\Deriv{\epsilon} m \deriv{\ell}m_2$ 
with $s_1 \approx m$ and $m_1\approx m_2$. By transitivity, $s_2 \approx m$, hence, by the stuttering property, 
$s_1$ is branching bisimilar to each place in the path from $s_2$ to $m$.
These constraints are not required by weak bisimilarity \cite{Mil89,Gor19b}:  given $s_1$ weak bisimilar to $s_2$, when matching a
transition $s_1 \deriv{\ell} m_1$ with $s_2 \Deriv{\epsilon} s_2' \deriv{\mu} m_2 \Deriv{\epsilon} m_2'$, 
weak bisimilarity only requires
that $m_1$ and $m_2'$ are weak bisimilar, but does not impose any condition on the intermediate states; in particular, it is not required that
$s_1$ is weak bisimilar to $ s_2'$, or that $m_1$ is weak bisimilar to $m_2$. 
\fine
\end{remark}

Now we provide a couple of examples showing that branching bisimulation equivalence is sensitive to the
timing of choice and to the kind of termination (i.e., to the size of the reachable marking).

\begin{figure}[t]
\centering

\begin{tikzpicture}[
every place/.style={draw,thick,inner sep=0pt,minimum size=6mm},
every transition/.style={draw,thick,inner sep=0pt,minimum size=4mm},
bend angle=45,
pre/.style={<-,shorten <=1pt,>=stealth,semithick},
post/.style={->,shorten >=1pt,>=stealth,semithick}
]
\def\eofigdist{3cm}
\def\eodist{0.3}
\def\eodisty{0.5}

\node (a) [label=left:$a)\qquad \qquad $]{};

\node (q1) [place] [label={above:$s_1$} ] {};
\node (t1) [transition] [below=\eodist of q1,label=left:$a\;$] {};
\node (q2) [place] [below=\eodist of t1,label=left:$s_2\;$] {};
\node (t2) [transition] [below left=\eodisty of q2,label=left:$b\;$] {};
\node (t3) [transition] [below right=\eodisty of q2, label=right:$\; c$] {};

\draw  [->] (q1) to (t1);
\draw  [->] (t1) to (q2);
\draw  [->] (q2) to (t2);
\draw  [->] (q2) to (t3);


\node (b) [right={2.2cm} of a, label=left:$b)\;\;$] {};

\node (p1) [place]  [right=\eofigdist of q1,label=above:$s_3$] {};
\node (s1) [transition] [below left=\eodisty of p1,label=left:$a\;$] {};
\node (s2) [transition] [below right=\eodisty of p1, label=right:$\; a$] {};

\node (p2) [place]  [below=\eodist of s1,label=left:$s_4 \;$] {};
\node (s3) [transition] [below=\eodist of p2,label=left:$b\;$] {};
\node (p3) [place]  [below=\eodist of s2,label=right:$\;s_5$] {};
\node (s4)  [transition] [below=\eodist of p3,label=right:$\;c$] {};

\draw  [->] (p1) to (s1);
\draw  [->] (p1) to (s2);
\draw  [->] (s1) to (p2);
\draw  [->] (s2) to (p3);
\draw  [->] (p2) to (s3);
\draw  [->] (p3) to (s4);


\node (c) [right={4.3cm} of b,label=left:$c)\;\;$] {};

\node (r1) [place]  [right=\eofigdist of p1,label=above:$s_6$] {};
\node (v1)  [transition] [below=\eodist of r1,label=right:$\;a$] {};
\node (r2) [place]  [below=\eodist of v1,label=below:$s_7$] {};

\node (r3) [place]  [right=\eodisty  of r1,label=above:$s_8$] {};
\node (v2)  [transition] [below=\eodist of r3,label=right:$\;a$] {};

\draw  [->] (r1) to (v1);
\draw  [->] (r3) to (v2);
\draw  [->] (v1) to (r2);

\end{tikzpicture}

\caption{Some non-branching bisimilar FSMs}
\label{abc-fig}
\end{figure}
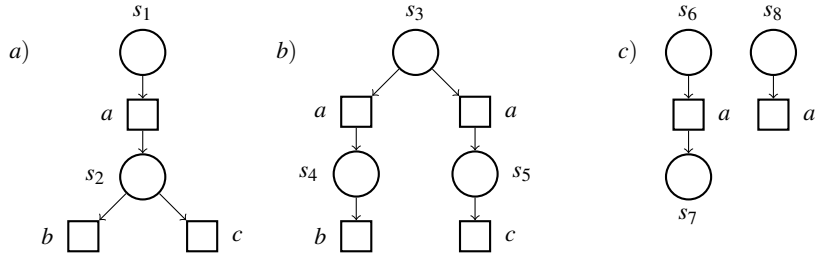

\begin{example}\label{anot-ex}
Consider Figure \ref{abc-fig}. It is not difficult to realize that $s_1 \not \approx s_3$.
In fact, $s_1$ may reach $s_2$ by performing $a$; $s_3$ can reply to this move in two different ways
by reaching either $s_4$ or $s_5$; however, while $s_2$ offers both $b$ and $c$, $s_4$
may perform only $b$ and $s_5$ only $c$; hence, $s_4$ and $s_5$ are not branching bisimilar to $s_2$ and so
also $s_1$ is not branching bisimilar to $s_3$. 
Moreover, also $s_6$ and $s_8$ are not branching bisimilar. In fact, $s_6$ can reach $s_7$ by performing $a$,
while $s_8$ can reply by reaching the empty marking, but $\theta \not\approx s_7$. This example 
shows that branching bisimilarity is sensitive to the kind of termination of a process: 
even if $s_7$ is stuck, it is not equivalent to
$\theta$ because the latter is the marking of a properly terminated process, while $s_7$ denotes a 
deadlock situation.
\fine
\end{example}

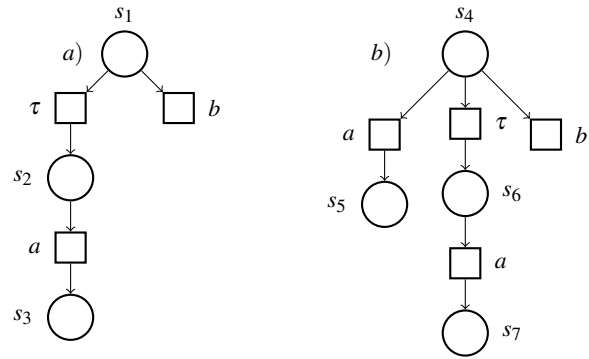
\begin{figure}[t]
\centering
\begin{tikzpicture}[
every place/.style={draw,thick,inner sep=0pt,minimum size=6mm},
every transition/.style={draw,thick,inner sep=0pt,minimum size=4mm},
bend angle=45,
pre/.style={<-,shorten <=1pt,>=stealth,semithick},
post/.style={->,shorten >=1pt,>=stealth,semithick}
]
\def\eofigdist{3.9cm}
\def\eodist{0.4}
\def\eodisty{0.9}
\def\eodistw{0.6}

\node (a) [label=left:$a) \quad $]{};

\node (q1) [place] [label={above:$s_1$} ] {};
\node (s1) [transition] [below left=\eodist of q1,label=left:$\tau\;$] {};
\node (s2) [transition] [below right=\eodist of q1,label=right:$\;b$] {};
\node (q2) [place] [below=\eodist of s1,label=left:$s_2\;$] {};
\node (s3) [transition] [below=\eodist of q2,label=left:$a\;$] {};
\node (q3) [place] [below=\eodist of s3,label=left:$s_3\;$] {};

\draw  [->] (q1) to (s1);
\draw  [->] (q1) to (s2);
\draw  [->] (s1) to (q2);
\draw  [->] (q2) to (s3);
\draw  [->] (s3) to (q3);

\node (b) [right={3.7cm} of a,label=left:$b)\;\;$] {};

\node (p1) [place]  [right=\eofigdist of q1,label=above:$s_4$] {};
\node (t1) [transition] [below left=\eodisty of p1,label=left:$a\;$] {};
\node (t2)  [transition] [below right=\eodisty of p1,label=right:$\;b$] {};
\node (t3)  [transition] [below =\eodist of p1,label=right:$\;\tau$] {};
\node (p2) [place] [below=\eodist of t3,label=right:$\;s_6$] {};
\node (t4)  [transition] [below =\eodist of p2,label=right:$\;a$] {};
\node (p3) [place] [below=\eodist of t1,label=left:$s_5\;$] {};
\node (p4) [place] [below=\eodist of t4,label=right:$\;s_7$] {};

\draw  [->] (p1) to (t1);
\draw  [->] (p1) to (t2);
\draw  [->] (p1) to (t3);
\draw  [->] (t3) to (p2);
\draw  [->] (p2) to (t4);
\draw  [->] (t1) to (p3);
\draw  [->] (t4) to (p4);
  
\end{tikzpicture}
\caption{Two non-branching bisimilar FSMs}
\label{br2-net}
\end{figure}

\begin{example}\label{br-ex21}
Consider the nets in Figure \ref{br2-net}. It is not difficult to see that $s_1$ and $s_4$ are weakly bisimilar \cite{Mil89,Gor19b}. However, 
$s_1 \not \approx s_4$, because to transition $s_4 \deriv{a} s_5$, place $s_1$ can only try to respond with
$s_1 \deriv{\tau} s_2 \deriv{a} s_3$, but not all the conditions required are satisfied; in particular,
$s_2 \not \approx s_4$, because only $s_4$ can do $b$. In fact, note that $s_1 \not \approx s_2$.
Indeed,
branching bisimulation equivalence ensures that, in response to the move $s_1 \deriv{\ell} m_1$, $s_2$ replies with 
$s_2 \Deriv{\epsilon} s \deriv{\ell} m_2$, in such a way that 
all the places traversed in the silent path from $s_2$ to $s$ are branching bisimilar.  
\fine
\end{example}

\begin{remark}\label{complexity1}{\bf (Complexity)}
If $m$ is the number of net transitions and $n$ of places, checking
whether two places of an FSM are branching bisimilar can be done in $O(m  \mbox{ log }n)$ time, 
by adapting the algorithm in \cite{JGKW20} for branching bisimulation on LTSs. 
\fine
\end{remark}

As branching bisimilarity is not a congruence for the CFM choice operator, we need a slightly more concrete
equivalence relation, which can be proved to be the coarsest congruence, contained in $\approx$, for all the CFM operators. 

\begin{definition}\label{def-rootbrbis}{\bf (Rooted branching bisimulation on places)}
Let $N = (S, A, T)$ be an FSM.
Two places $s_1$ and $s_2$ are {\em rooted branching bisimilar}, denoted $s_1 \approx_{c} s_2$,  if $\forall \ell \in A$
\begin{itemize}
\item for all $m_1$ such that  $s_1\deriv{\ell}m_1$, there exists $m_2$ such that $s_2\deriv{\ell}m_2$ and 
either $m_1 = \theta = m_2$ or $m_1 \approx m_2$,
\item for all $m_2$ such that $s_2\deriv{\ell}m_2$, there exists $m_1$ such that $s_1\deriv{\ell}m_1$ and 
either $m_1 = \theta = m_2$ or $m_1 \approx m_2$.\\[-1cm]
\end{itemize}
\fine
\end{definition}

The peculiar feature of $\approx_{c}$ is that initial moves are matched as in strong bisimulation \cite{Mil89}, 
while subsequent moves are matched as for branching bisimilarity.
Therefore, rooted branching bisimilarity is a slightly finer variant of branching bisimilarity.

\begin{proposition}\label{bis-br-eq}\cite{Gor19b}
For each FSM $N = (S, A, T)$ with silent moves, relations $\approx$ and  $\approx_c$ are equivalence relations.
\fine
\end{proposition}

%
\subsection{(Rooted) Branching Team Equivalence} \label{hteam-eq-sec}
%

\begin{definition}\label{hadd-eq}{\bf (Additive closure)}
Given an FSM net $N = (S, A, T)$ and a {\em place relation} $R \subseteq S \times S$, we define 
a {\em marking relation}
$R^\oplus \, \subseteq \, {\mathcal M}(S) \times {\mathcal M}(S)$, called 
the {\em additive closure} of $R$,
as the least relation induced by the following axiom and rule.

$\begin{array}{lllllllllll}
 \bigfrac{}{(\theta, \theta) \in  R^\oplus} & \; \; \; & \; \; \; 
 \bigfrac{(s_1, s_2) \in R \; \; \; (m_1, m_2) \in R^\oplus }{(s_1 \oplus m_1, s_2 \oplus m_2) \in  R^\oplus }  \\
\end{array}$
\\[-.2cm]
\fine
\end{definition}

Note that, by definition, two markings are related by $R^\oplus$ only if they have the same size.
An alternative way to define that two markings $m_1$ and $m_2$
are related by $R^\oplus$ is to state that $m_1$ can be represented as 
$s_1 \oplus s_2 \oplus \ldots \oplus s_k$, $m_2$ can be represented as 
$s_1' \oplus s_2' \oplus \ldots \oplus s_k'$ and $(s_i, s_i') \in R$ for $i = 1, \ldots, k$.

\begin{remark}{\bf (Additivity and subtractivity)}\label{add-sub-rem}
The additive closure $R^\oplus$ of a place relation $R$ is {\em additive}: 
if $(m_1, m_2) \in R^\oplus$ and $(m_1', m_2') \in R^\oplus$,
then $(m_1 \oplus m_1', m_2 \oplus m_2') \in R^\oplus$. Moreover,
if $R$ is an equivalence relation, then $R^\oplus$ is {\em subtractive}: 
if $(m_1 \oplus m_1',$ $ m_2 \oplus m_2') \in  R^\oplus$ 
and $(m_1, m_2) \in R^\oplus$,
then $(m_1', m_2') \in R^\oplus$.
(Proof in \cite{Gor17b}).
\fine
\end{remark}

\begin{definition}\label{bteam-eq}{\bf (Rooted branching team equivalence)}
Given a finite-state machine $N = (S, A, T)$ and branching bisimulation equivalence $\approx$ and rooted branching bisimulation equivalence
${\approx_c}$, we define  {\em  branching team/rooted branching team equivalences} as the relations
$\approx^\oplus$ and $\approx_c^\oplus$, respectively.
\fine
\end{definition}

Of course, (rooted) branching team equivalence relates markings of the same size only;
moreover, $\approx^\oplus$ and $\approx_c^\oplus$ are equivalence relations by Proposition \ref{bis-br-eq},
as the additive closure of an equivalence relation is also an equivalence relation.

Moreover, it is easy to prove that $\approx^\oplus$ enjoys a form of stuttering property (cf. Remark \ref{stutt-prop-rem}), concerning
{\em sequential} silent transition sequences:

if $m_0[t_1\rangle m_1 [t_2\rangle \ldots m_{n-1} [t_n\rangle m_n$, $l(t_i) = \tau$ for $i = 1, \ldots, n$ 
(hence, the transition sequence $t_1 t_2 \ldots t_n$ is silent), and $\pre{t_{i}} = \post{t_{i-1}}$
for $i = 2, \ldots, n$ (hence, the transition sequence $t_1 t_2 \ldots t_n$ is sequential), and, 
moreover, $m_0 \approx^\oplus m_n$, 
then
$m_i \approx^\oplus m_j$ for $i,j = 1, \ldots, n$.

In \cite{Gor19b} we proved that whenever $m_1 \approx^\oplus m_2$, if $m_1[t_1\rangle m_1'$, then
 \begin{itemize}
    \item[--]  either $l(t_1) = \tau$ and 

       \begin{itemize}
      \item[(i)] either $\exists \sigma_2$ nonempty, silent, sequential, such that
      $\pre{t_1} \approx \pre{\sigma_2}$, 
      $\post{t_1} \approx \post{\sigma_2}$, 
       $\pre{t_1} \approx \post{\sigma_2}$,
      $m_2[\sigma_2\rangle m_2'$ with $m_1 \approx^\oplus m_2'$ and $m_1' \approx^\oplus m_2'$,
    
     \item[(ii)] or $\exists s_2 \in m_2$ such that $\pre{t_1} \approx s_2$,  $\post{t_1} \approx s_2$, 
     with  $m_1' \approx^\oplus m_2$,
     \end{itemize}
     
   \item[--] or $\exists \sigma,  t_2$ such that $\sigma t_2$ is sequential, $\sigma$ is silent, $l(t_1) = l(t_2)$, 
     $\pre{t_1} \approx \pre{\sigma t_2}$,  $\pre{t_1} \approx \pre{t_2}$, $\post{t_1} \approx^\oplus \post{t_2}$,
     $m_2[\sigma\rangle m [t_2\rangle m_2'$ with $m_1 \approx^\oplus m$ and $m_1' \approx^\oplus m_2'$;
   \end{itemize}
and, symmetrically, $\forall t_2$ such that  $m_2[t_2 \rangle m_2'$.
In the either-(i)-case, we can easily prove that all the markings in the path from $m_2$ to $m_2'$ 
are branching team bisimilar;
similarly, in the or-case, all the markings in the path from $m_2$ to $m$ are branching team bisimilar. Hence,
branching team bisimilarity $\approx^\oplus$ does fully respect the timing of choices.

\begin{remark}\label{complexity-team}{\bf (Complexity)}
Once the place relations $\approx$ and $\approx_c$ have been computed once and for all for the given net 
in $O(m \cdot \mbox{ log }n)$ time, the algorithm in \cite{Gor17b} checks whether 
two markings $m_1$ and $m_2$ of equal size $k$ 
are team equivalent in $O(k^2)$ time. 
\fine
\end{remark}

%
\section{CFM: Syntax, Semantics, Axiomatization} \label{cfm-sec}
%

\subsection{Syntax}\label{cfm-syntax}

Let $Act = H \cup L \cup \{\tau\}$ be a finite set of actions, ranged over by $\mu$, composed of two disjoint
subsets $L$ and $H$ of low-level actions and high-level ones, respectively, and by the silent action $\tau$.
Let $\cons$ be a finite set of constants, disjoint from 
$Act$, ranged over by $A, B, C,\ldots$. 
The CFM {\em terms} (where CFM is the acronym of {\em Concurrent Finite-state Machines}) 
are generated from actions and constants by the 
following abstract syntax (with three syntactic categories):

$\begin{array}{lccccccccccl}
s &  ::= &  \nil & | & \mu.q & | &  s+s & \hspace{1 cm} \mbox{{\em guarded processes}}\\
q & ::= & s & | & C & &&\hspace{1 cm} \mbox{{\em sequential processes}}\\
p & ::= & q & | & p \para p & & &\hspace{1 cm} \mbox{{\em parallel processes}}\\
\end{array}$

\noindent
where $\nil$ is the empty process, $\mu.q$ is a process where
action $\mu$ prefixes the residual $q$ ($\mu.-$ is the {\em action prefixing} operator),
$s_1 + s_2$ denotes the alternative composition of $s_1$ and $s_2$ ($- + -$ is
the {\em choice} operator), $p_1 \para p_2$ denotes the asynchronous parallel composition of $p_1$ and $p_2$
and $C$ is a constant. 
A constant $C$ may be equipped with a definition, but this must be a guarded process, i.e., 
$C \eqdef s$. 
A term $p$ is a CFM {\em process} if each constant in \const{p} (the set of constants used by $p$; see \cite{Gor17} for details) is equipped with a defining equation (in category $s$).
The set of CFM processes is denoted by $\mathcal{P}_{CFM}$, the set of its sequential processes, i.e.,
those in syntactic category $q$, by $\mathcal{P}_{CFM}^{seq}$ and  the set of its guarded processes, i.e.,
those in syntactic category $s$, by $\mathcal{P}_{CFM}^{grd}$.
By $sort(p) \subseteq Act$ we denote the set of all the actions occurring in $p$ and in the body of the constants in \const{p}.

\begin{table}[t]
{\renewcommand{\arraystretch}{1.3}
\hrulefill\\[-0.8cm]

\begin{center}
$\begin{array}{lcllcl}
\mbox{(Pref)}  &\bigfrac{}{\mu.p\deriv{\mu}p} & \qquad & \; \; \;
\mbox{(Cons)} & \bigfrac{p\deriv{\mu}p'}{C\deriv{\mu}p'}& C \eqdef p \\
\mbox{(Sum$_1$)}  & \bigfrac{p\deriv{\mu}p'}{p+q\deriv{\mu}p'} &  \qquad & \; \; \;
\mbox{(Sum$_2$)}  & \bigfrac{q\deriv{\mu}q'}{p+q\deriv{\mu}q'}\\
\mbox{(Par$_1$)}  & \bigfrac{p\deriv{\mu}p'}{p\para q\deriv{\mu}p'\para q} & & \; \; \;
\mbox{(Par$_2$)}  & \bigfrac{q\deriv{\mu}q'}{p\para q\deriv{\mu}p \para q'} \\[-.2cm]
\end{array}$
\end{center}}
\hrulefill
\caption{Structural operational LTS semantics for CFM}\label{cfm-lts}
\end{table}

\begin{table}[!t]

{\renewcommand{\arraystretch}{1.2}
\hrulefill\\[-.7cm]
\begin{center}
$\begin{array}{rcllrcllllll}
\dec(\nil) & = & \theta & \qquad &
\dec(\mu.p) & = &  \{ \mu.p\}  \\  
\dec(p + p') & = &   \{p + p'\} & \qquad &
\dec (C) & = & \{C\}\\
\dec(p \para p') & = &   \dec(p) \oplus \dec(p')\\[-.2cm]
\end{array}$
\end{center}}
\hrulefill
\caption{Decomposition function} \protect\label{dec1}
\end{table}

\subsection{Semantics}\label{net-sem-sec}

The interleaving LTS semantics for CFM is given by the structural operational rules in Table \ref{cfm-lts}. 
Note that each state
of the LTS is a CFM process. As an example, the LTS for $C \para B$, with $C \eqdef h.B$ 
and $B \eqdef l.B$, is described in
Figure \ref{prima-fig}(c). It is possible to prove that, for any $p$, the set of states reachable 
from $p$ is finite \cite{Gor17}.

The Petri net semantics for CFM  is such that the set $S_{CFM}$ of places  is the set of 
the sequential CFM processes, 
without $\nil$, i.e., $S_{CFM} = {\mathcal P}_{CFM}^{seq} \setminus \{\nil\}$.
The decomposition function $\dec: {\mathcal P}_{CFM}  \rightarrow {\mathcal M}(S_{CFM})$, mapping process terms to markings,
is defined in Table \ref{dec1}.
An easy induction proves that for any $p \in \mathcal{P}_{CFM}$, $\dec(p)$ is a finite multiset of sequential processes. 
Note that, if $C \eqdef \nil$, then $\theta = \dec(\nil) \neq \dec(C) = \{ C \}$. Note also that $\theta = \dec(\nil) \neq \dec(\nil + \nil) = \{\nil + \nil\}$, which is a deadlock place.

\begin{table}[!t]
{\renewcommand{\arraystretch}{1.2}
\hrulefill\\[-1.1cm]

\begin{center}
$\begin{array}{rcllcllcl}
 \encodings{I}{\nil}  & =  & (\emptyset, \emptyset, \emptyset, \theta) &\\
 \encodings{I}{\mu.p}  & =  & (S, A, T, \{\mu.p\}) & \mbox{ given } 
  \encodings{I}{p}   =   (S', A', T', \dec(p)) \; \mbox{ and where } \\ & & &  S = \{\mu.p\} \cup S', \; A = \{\mu\} \cup A', \; 
  T = \{(\{\mu.p\}, \mu, \dec(p))\} \cup T'\\
 \encodings{I}{p_1 + p_2}  & =  & (S, A, T, \{p_1 + p_2\})& \mbox{ given }  \encodings{I}{p_i}   =   
 (S_i, A_i, T_i, \dec(p_i)) \; \mbox{ for $i = 1, 2$, and where} \\
  & & &  
  S = \{p_1 + p_2\} \cup S_1' \cup S_2', \mbox{ with, for $i = 1, 2$, } \\
  & & & S'_i = \begin{cases}
   S_i & \! \! \mbox{
   $\exists t \in T_i$ such that $\post{t}(p_i) > 0$}\\ 
  S_i \setminus \{p_i\} & \! \! \mbox{otherwise} \\
   \end{cases}\\
& & &  A = A_1 \cup A_2, \;  T = T' \cup T'_1 \cup T'_2,  \mbox{ with, for $i = 1, 2$, }\\
&&& T'_i = \begin{cases}
   T_i  & \! \! \mbox{
   $\exists t \in T_i \,.\, \post{t}(p_i) > 0$}\\ 
 T_i \setminus \{t \in T_i \mid \pre{t}(p_i) > 0\}  & \! \! \mbox{otherwise} \\
   \end{cases}\\ & & & 
  T' = \{(\{p_1 + p_2\}, \mu, m) \mid (\{p_i\}, \mu, m) \in T_i, i = 1, 2\}\\
 \encodings{I}{C}  & =  & (\{C\}, \emptyset, \emptyset, \{C\}) & \mbox{ if $C \in I$ } \\
  \encodings{I}{C}  & =  & (S, A, T, \{C\}) & \mbox{ if $C \not \in I$, given $C \eqdef p$ and } 
  \encodings{I\cup \{C\}}{p}   =   (S', A', T', \dec(p)) \\\ 
&&&  A = A', S =  \{C\} \cup S'', \mbox{ where }\\
  & & & S'' =  \begin{cases}
   S'  & \! \! \mbox{
   $\exists t \in T' \, .\,\post{t}(p) > 0$}\\ 
   S' \setminus \{p\} & \! \! \mbox{otherwise}\\ 
   \end{cases}\\
  & & & T = \{(\{C\}, \mu, m) \mid (\{p\}, \mu, m) \in T'\} \cup T'' \mbox{ where }\\ & & &
  T'' =   \begin{cases}
   T'  & \! \! \mbox{
   $\exists t \in T' \, . \,\post{t}(p) > 0$}\\ 
   T' \setminus \{t \in T' \mid \pre{t}(p) > 0\} & \! \! \mbox{otherwise} \\
   \end{cases}\\ 
   \encodings{I}{p_1 \para p_2}  & =  & (S, A, T, m_0) & \mbox{ given } 
 \encodings{I}{p_i}   =   (S_i, A_i, T_i, m_i) \; \mbox{ for $i = 1, 2$, and where } \\ & & &  
  S = S_1 \cup S_2, \;  A = A_1 \cup A_2, \;  T = T_1 \cup T_2, \; m_0 = m_1 \oplus m_2\\[-0.2cm]
\end{array}$
\end{center}}
\hrulefill
\caption{Denotational net semantics}\label{den-net-cfm}
\end{table}

The net system $\encodings{\emptyset}{p}$
associated with process $p$ is defined in a denotational style. The details of
the construction are outlined in Table \ref{den-net-cfm}. The mapping is parametrized 
by a set of constants
that have  already been found while scanning $p$; such a set is initially empty and it is used to avoid 
looping on recursive constants. The definition is syntax driven
and also the places of the constructed net are syntactic objects, i.e., CFM sequential process terms.
E.g., the net system $\encodings{\emptyset}{a.\nil}$ is a net composed of one single marked place, 
namely process $a.\nil$, and one single transition $(\{a.\nil\}, a, \theta)$. 
A bit of care is needed in the rule for choice: in order to include only strictly necessary 
places and transitions, the initial place $p_1$ (or $p_2$) of the subnet $\encodings{I}{p_1}$ (or $\encodings{I}{p_2}$) 
is to be kept in the net for $p_1 + p_2$ only if there exists a transition reaching place $p_1$ (or $p_2$) in $\encodings{I}{p_1}$
(or $\encodings{I}{p_2}$), 
otherwise $p_1$ (or $p_2$) can be safely removed in the new net.
Similarly, for the rule for constants.
Examples of the net construction for some CFM terms can be found in \cite{Gor17,Gor19b}. 

This net semantics has the following important properties:
\begin{itemize}
\item the semantics of a CFM process term $p$, i.e., the net  $\encodings{\emptyset}{p}$, is a finite-state machine, 
whose initial marking is $\dec(p)$; moreover,
\item for any finite-state machine $N(m_0)$ (whose places and transitions are all reachable from $m_0$), there exists a CFM process term $p_{N(m_0)}$ such that its semantics
$\encodings{\emptyset}{p_{N(m_0)}}$ is a net isomorphic to $N(m_0)$ ({\em Representability Theorem}).
\end{itemize}
Therefore, thanks to these results (proved in \cite{Gor17}), we can conclude that the CFM process algebra truly 
represents the class of finite-state machines, up to isomorphism.
Hence, we can transpose the definition of (rooted) branching team equivalence 
from finite-state machines to CFM process terms in a simple way.

\begin{definition}\label{def-cfm-hteam}
Two CFM processes $p$ and $q$ are (rooted) branching team equivalent, 
denoted $p \approx^\oplus q$ and $p \approx_c^\oplus q$, respectively, 
if, by taking the 
(union of the) nets $\encodings{\emptyset}{p}$
and $\encodings{\emptyset}{q}$, we have that  $\dec(p) \approx^\oplus \dec(q)$ and $\dec(p) \approx_c^\oplus \dec(q)$, respectively.
\fine
\end{definition}

Of course, for sequential CFM processes, (rooted) branching team equivalence $\approx^\oplus$ ($\approx_c^\oplus$) coincides with 
(rooted) branching bisimilarity on places $\approx$ ($\approx_c$). 

Finally, as we are going to use an auxiliary restriction operator over CFM terms of the form $p \setminus H$, we define
its net semantics as follows.

\begin{definition}\label{restr-cfm}{\bf (Semantics of the auxiliary restriction operator)}
Given a CFM process $p$, whose net semantics is  $\encodings{\emptyset}{p}   =   (S, A, T, \dec(p))$,
we define the net associated to $p \setminus H$ as the net $\encodings{\emptyset}{p\setminus H}   =   (S', A', T', m)$ where
\begin{itemize}
\item $S' = \{s\setminus H \mid s \in S\}$, i.e., each place is decorated by the restriction operator;
\item $A' = A \setminus H$, i.e., $\{\mu \mid \mu \in A, \mu \not\in H\}$;
\item $T' = \{(\pre{t}\setminus H, l(t), \post{t}\setminus H) \mid t \in T, l(t) \not\in H\}$;
\item $ m = \dec(p) \setminus H$, where the restriction operator is applied element-wise to the places, if any, of the 
marking $\dec(p)$.\\[-.9cm]
\end{itemize}
\fine
\end{definition}

As an example, the net for $C\para B$, with $C \eqdef h.B$ and $B \eqdef l.B$, is outlined in Figure \ref{seconda-fig}(a), while, assuming that $H$ is composed of the single action $h$, the net for $(C \para B) \setminus h$ is in Figure \ref{seconda-fig}(b).

Branching team bisimilarity is a congruence  for action prefixing, parallel composition, recursion (via process constants) 
and restriction, but not for the choice operator. For instance,
$\tau.a.\nil \approx^\oplus a.\nil$, but $b.\nil + \tau.a.\nil \not\approx^\oplus b.\nil + a.\nil$. However, rooted branching team 
bisimilarity $\approx_c^\oplus$ is a congruence also for
the choice operator, so that it can be axiomatized. (More details on congruence properties and algebraic properties in \cite{Gor19b}.)

\subsection{Axiomatization}

In this section we recall the sound and complete, finite axiomatization of rooted branching team equivalence $\approx_c^\oplus$ over CFM, outlined in \cite{Gor19b}, where the reader can find more detail.
For simplicity's sake, the syntactic definition of {\em open} CFM (i.e., CFM with variables) is given with only one syntactic category, but 
each ground instantiation of an axiom must respect the syntactic definition of CFM given (by means of three syntactic categories) in 
Section \ref{cfm-syntax}; this means that we can write the axiom $x + (y + z) = (x + y) + z$, but 
it is invalid to instantiate it to $C + (a.\nil + b.\nil) = (C + a.\nil) + b.\nil$ because these are not legal CFM 
processes (the constant $C$ cannot be used as a summand).

\begin{table}[t]
{\renewcommand{\arraystretch}{1.1}
\hrulefill\\[-.7cm]
\normalsize{

\begin{center}
$\begin{array}{llrcll}
{\bf A1} &\; \;  \mbox{Associativity} &\; \;  x + (y + z) & = & (x + y) + z &\\
{\bf A2} &\; \;  \mbox{Commutativity} &\; \;  x + y & = & y + x& \\
{\bf A3} &\; \;  \mbox{Identity} &\; \;  x + \nil & = & x & \quad  \mbox{ if $x \neq \nil$}\\
{\bf A4} &\; \;  \mbox{Idempotence} &\; \;  x + x & = & x & \quad  \mbox{ if $x \neq \nil$}\\
\end{array}$

\hrulefill

$\begin{array}{llrcll}
{\bf B} &\; \;  \mbox{} &\; \;  \mu.(\tau.(x + y) + x) & = & \mu.(x + y)  & \\
\end{array}$

\hrulefill

$\begin{array}{llrcllll}
{\bf R1} & \;  \mbox{Stuck} & \mbox{if $C \eqdef \nil$,} & \mbox{then} &
\mbox{$C = \nil + \nil$} &\\
{\bf R2} & \;  \mbox{Unfolding} &  \mbox{if $C \eqdef p \; \wedge \;  p \neq \nil$,} & \mbox{then} &
\mbox{$C = p$} &\\
{\bf R3} & \;  \mbox{Folding} &  \mbox{if $C \eqdef p\{C/x\}  \; \wedge \; og(p) \; \wedge \;
q = p\{q/x\}$,} & \mbox{then} & \mbox{$C = q$} & \\
\end{array}$

\hrulefill

$\begin{array}{llrlllll}
{\bf U1} &\;    \mbox{if $C \eqdef  (\tau.x + p)\{C/x\} \wedge D\eqdef ( \tau.(p + \nil) + p)\{D/x\}$} & \mbox{then} &
C  =  D &\\
{\bf U2} & \;    \mbox{if $C \eqdef (\tau.(\tau.x + p) + r)\{C/x\} \wedge D \eqdef (\tau.(p + r) + r)\{D/x\}$} & \mbox{then} & 
C  =  D & \\
\end{array}$

$\begin{array}{llrcllll}
{\bf U3} & \;  \mbox{if $C \eqdef (\tau.(\tau.q + p) + r)\{C/x\} \wedge  D   \eqdef  (\tau.(q + p) + r)\{D/x\}$,}\\
& \mbox{ $x$ unguarded in $q\in  \mathcal{P}_{CFM}^{grd}$,} & \mbox{ then} 
&    C   =   D\\
{\bf U4} & \;  \mbox{if $C \eqdef (\tau.(\tau.x + p) + \tau.(\tau.x + q) + r)\{C/x\}$}\\
& \mbox{$\wedge \; D \eqdef (\tau.(\tau.x + p + q) + r)\{D/x\}$}
 &  \mbox{ then} 
&    C  =   D\\
\end{array}$

\hrulefill

$\begin{array}{llrcll}
{\bf P1} &\; \;  \mbox{Associativity} &\; \;  x \para (y \para z) & = & (x \para y) \para z &\\
{\bf P2} &\; \;  \mbox{Commutativity} &\; \;  x \para y & = & y \para x & \\
{\bf P3} &\; \;  \mbox{Identity} &\; \;  x \para \nil & = & x &\\[-0.3cm]
\end{array}$
\end{center}}
\hrulefill
}
\caption{Axioms for rooted branching team equivalence}\label{axiom-tab}
\end{table}

The set of axioms are outlined in Table \ref{axiom-tab}. 
We call $E$ the set of axioms $\{${\bf A1, A2, A3, A4, B, R1, R2, R3, U1, U2, U3, U4, P1, P2, P3}$\}$. 
By the notation $E \vdash p = q$ we mean that there exists an equational deduction proof 
of the equality $p = q$, by using the axioms in $E$. Besides the usual equational deduction rules of reflexivity, symmetry, transitivity,
substitutivity and instantiation (see, e.g., \cite{GV15}), in order to deal with constants we need also the following recursion
congruence rule:
\[
\frac{p = q \; \wedge \; A \eqdef p\{A/x\} \; \wedge \; B \eqdef q\{B/x\}}{A = B}
\]
\noindent 
where $p\{A/x\}$ denotes the open term $p$ where all occurrences of the variable $x$ are replaced by $A$.
The axioms {\bf A1-A4} are the usual axioms for choice where, however, {\bf A3-A4} have the side 
condition $x \neq \nil$; hence, it is not possible to prove $E \vdash \nil + \nil = \nil$, as expected, because 
these two terms have a completely different semantics. Axiom {\bf B}, originally in \cite{vGW96}, is the axiom for rooted branching bisimilarity.
The conditional axioms {\bf R1-R3} are about process constants. Note that  {\bf R2}
requires that $p$ is not (equal to) $\nil$ (condition $p \neq \nil$). 
Note also that these conditional axioms are actually a finite collection of axioms, one for each constant 
definition: since the set $\cons$ of process constants is finite, the instances of {\bf R1-R3} are finitely many. 
Note that axiom {\bf R3} requires that the body $p$ of the constant $C \eqdef p\{C/x\}$ is {\em observationally guarded} 
(condition $og(p)$), meaning that
 $C \NDeriv{\tau} C$.
 The axioms  $\{${\bf U1, U2, U3, U4}$\}$ are necessary to prove (soundness and) completeness
for the whole of CFM, 
hence including observationally unguarded (i.e., possibly silently divergent) processes.
Finally, we have axioms {\bf P1-P3} for parallel composition. 

\begin{theorem}\cite{Gor19b}{\bf (Sound and Complete)}\label{sound-complete-th}
For every $p, q \in  \mathcal{P}_{CFM}$,  $E \vdash p = q$ if and only if $p \approx_c^\oplus q$.
\fine
\end{theorem}

%
\section{DNI: Distributed Non-Interference} \label{dni-sec}
%

\subsection{Definition and Compositional Verification}

\begin{definition}\label{def-dni}{\bf (Distributed Non-Interference (DNI))}
A CFM process $p$ enjoys the distributed non-interference property (DNI, for short) if
for each $p', p''$, reachable from $p$, and for each $h \in H$,
such that $p' \deriv{h} p''$, we have that  $p' \setminus H \approx^\oplus p''\setminus H$ holds.
\fine
\end{definition}

This intuitive and simple definition is somehow hybrid, because, on the one hand, it refers to reachable states $p'$ and $p''$ in the LTS
semantics, while, on the other hand, it requires that $p' \setminus H \approx^\oplus p''\setminus H$, a condition that can be checked on the
Petri net semantics. We can reformulate this definition in such a way that it refers only to the Petri net semantics, 
a reformulation that will be very useful in proving the following theorem.

\begin{definition}\label{def-dni}{\bf (DNI on Petri nets)}
Given a CFM process $p$ and the FSM $\encodings{\emptyset}{p}$ and its low-observable subnet
$\encodings{\emptyset}{p\setminus H}$,
we say that $p$ satisfies DNI if for each $m_1, m_2$, reachable from $\dec(p)$ in $\encodings{\emptyset}{p}$, and for 
each $h \in H$, such that $m_1 \deriv{h} m_2$, we have that  the two markings $m_1 \setminus H$ and $m_2 \setminus H$ of
$\encodings{\emptyset}{p\setminus H}$ are branching team equivalent.
\fine
\end{definition}

\begin{theorem}\label{th-dni}
A  process $p$ is not DNI if and only if there exists $p_i \in \dec(p)$ such that $p_i$ is not DNI.
\proof ($\Rightarrow$) If $p$ is not DNI, then there exist $m_1, m_2$ reachable from $\dec(p)$, and a high-level action $h \in H$, such that $m_1 \deriv{h} m_2$,
but the two markings $m_1 \setminus H$ and $m_2 \setminus H$ of
$\encodings{\emptyset}{p\setminus H}$ are not branching team equivalent. Because of the shape of FSM transitions, $m_1 \deriv{h} m_2$
is a move that must be due to a net transition $s \deriv{h} m$, so that $m_1 = s \oplus \overline{m}$ and $m_2 = m \oplus \overline{m}$.
Hence, $m_1 \setminus H =$ $ s \setminus H \oplus \overline{m}\setminus H$ and $m_2 \setminus H = m \setminus H 
\oplus \overline{m}\setminus H$. If $m_1 \setminus H$ is not branching team equivalent to $m_2 \setminus H$, then necessarily
$s \setminus H \not \approx m \setminus H$, because $\overline{m}\setminus H$ is certainly branching team equivalent to itself.
Since $m_1$ is reachable from $\dec(p)$, because of the shape of net transitions, there exists $p_i \in \dec(p)$ such that $s$ is reachable 
from $p_i$. Summing up, if $p$ is not DNI, then we have found a sequential subprocess $p_i \in \dec(p)$ which is not DNI, because
$p_i$ can reach $s$, transition $s \deriv{h} m$ is executable and $s \setminus H \not \approx m \setminus H$. 

($\Leftarrow$) The converse implication is obvious.
\fine
\end{theorem}

\begin{corollary}\label{cor-dni1}
A CFM process $p$ is DNI if and only if each $p_i \in dom(\dec(p))$ is  DNI.
\proof The thesis is just the contranominal of Theorem \ref{th-dni}.
\fine
\end{corollary}

Hence, in order to check whether $p$ is DNI, we first compute $\dec(p)$ to single out its sequential components;
then, we consider only the elements of $dom(\dec(p))$, because it is necessary to check each sequential component
only once. For instance, if $p = (q_1 \para q_2) \para (q_1\para q_2)$, then, assuming $q_1$ and $q_2$ sequential,
$\dec(p) = 2 \cdot q_1 \oplus 2 \cdot q_2$, so that $dom(\dec(p)) = \{q_1, q_2\}$,
 and so we have simply to check whether $q_1$ and $q_2$ are DNI.

\begin{corollary}\label{cor-dni2}
If $p \approx^\oplus q$ and $p$ is DNI, then also $q$ is DNI.
\proof By Corollary \ref{cor-dni1}, $p$ is DNI if and only if each $p_i \in dom(\dec(p))$ is  DNI.
Since $p \approx^\oplus q$, there exists a $\approx$-relating bijection between $\dec(p)$ and $\dec(q)$.
Therefore, the thesis is implied by the following obvious fact: given two sequential CFM processes $p_i, q_j$ such that
$p_i \approx q_j$, if $p_i$ is DNI then $q_j$ is DNI. In fact, if $p_i$ is DNI, then for all $s_i$ reachable from $p_i$,
if $s_i \deriv{h} m_i$, then $s_i \setminus H \approx m_i \setminus H$. Since $p_i \approx q_j$, for each such $s_i$ there exists a
suitable $s_j$ reachable from $p_j$ such that 
$s_i \approx s_j$ and so also $s_i \setminus H \approx s_j \setminus H$ by congruence w.r.t. restriction. 
Hence, since $s_i \deriv{h} m_i$, there exists $m_j$ such that $s_j \deriv{h} m_j$ with $m_i \approx m_j$
and so also $m_i \setminus H \approx m_j \setminus H$ by congruence. It follows trivially by transitivity 
that $s_j \setminus H \approx m_j \setminus H$, as required. 

The case when $q_j$ moves first is symmetric: for each $s_j$ reachable from $q_j$, since $p_i \approx q_j$, 
there exists a
suitable $s_i$ reachable from $p_i$ such that 
$s_i \approx s_j$ and so also $s_i \setminus H \approx s_j \setminus H$ by congruence w.r.t. restriction.
If $s_j \deriv{h} m_j$, then, since $s_i \approx s_j$,
also $s_i \deriv{h} m_i$ with $m_i \approx m_j$ and so, by congruence w.r.t. restriction, also $m_i \setminus H \approx m_j \setminus H$. 
Since $p_i$ is DNI, we get $s_i \setminus H \approx m_i \setminus H$. The thesis then follows by transitivity:
$s_j \setminus H \approx m_j \setminus H$.
\fine
\end{corollary}

\subsection{Efficient Verification Based on a Structural Characterization}

Because of the shape of finite-state machines and because of additivity (and subtractivity) of the behavioral relation $\approx^\oplus$
(cf. Remark \ref{add-sub-rem}), 
a very efficient DNI verification can be done by the following polynomial algorithm. 

Given the CFM process $p$, first compute
the nets $\encodings{\emptyset}{p} = (S, A, T, \dec(p))$ and the low-observable subnet $\encodings{\emptyset}{p\setminus H}$. Then, compute 
branching bisimilarity $\approx$ on the places of the net $\encodings{\emptyset}{p\setminus H}$. 
Finally, for each $t \in T$ such that $l(t) \in H$,
check whether $\pre{t}\setminus H$ and $\post{t}\setminus H$ are branching bisimilar in the net $\encodings{\emptyset}{p\setminus H}$: 
if this is the case for all the high-level transitions of 
$\encodings{\emptyset}{p}$, then $p$ is DNI; on the contrary, if for some $t$ the check fails (e.g., because $\post{t} = \theta$), 
then $p$ is not DNI.

The correctness of this polynomial algorithm follows by the fact that the net $\encodings{\emptyset}{p}$ is {\em dynamically 
reduced}, i.e., all the places and the transitions are reachable from the initial marking $\dec(p)$. 
Once $\encodings{\emptyset}{p}$ has been computed by the polynomial algorithm described in
Table \ref{den-net-cfm}, and $\encodings{\emptyset}{p\setminus H}$ is computed by the simple algorithm in Definition \ref{restr-cfm},
the complexity of this procedure is essentially related to the problem of computing  $\approx$ (cf. Remark \ref{complexity1}) for 
$\encodings{\emptyset}{p\setminus H}$ and to give it a suitable adjacency 
matrix representation in order to check easily, for each high-level transition $t\in T$, whether the
two relevant places $\pre{t}\setminus H$ and $\post{t}\setminus H$ are related by $\approx$.

\subsection{Typing System}

In this section we provide a syntactic characterization on CFM of a slight strengthening of DNI, which is based on {\em rooted} 
branching team equivalence $\approx_c$, rather than on $\approx$.
With abuse of notation, also this property is called DNI: $p$ is DNI if
for each $p'$, reachable from $p$, and for each $h \in H$,
such that $p' \deriv{h} p''$ for some $p''$, we have that  $p' \setminus H \approx_c^\oplus p''\setminus H$ holds.

This syntactic characterization is based on a typing proof system, which 
exploits the axiomatization of rooted branching team equivalence $\approx_c^\oplus$ in Table \ref{axiom-tab}.
Let us first define an auxiliary operator $r(-)$, which takes in input a CFM process $p$ and returns a CFM 
process $p'$ obtained from $p$ by pruning its high-level actions, so that $sort(p') \subseteq L \cup \{\tau\}$. Its definition
is outlined in Table \ref{restr}, where $r(h.p) = \nil + \nil$ because the pruning of $h.p$ is to be considered 
as a deadlock place.
For instance, consider $C \eqdef h.l.C  + l.C$; then $r(C) = C'$, where
$C' \eqdef \nil + \nil + l.C'$. Similarly, if $D \eqdef l.h.D$, then $r(D) = D'$ where $D' \eqdef l.(\nil+\nil)$.
It is a trivial observation that the net semantics of $r(p)$ is isomorphic to the (reachable part of the) 
net semantics of $p \setminus H$ for any CFM process $p$.
Moreover, we also define the set of initial actions of $p$, denoted by $In(p)$, whose definition is outlined in
Table \ref{initial}.

\begin{table}[t]

{\renewcommand{\arraystretch}{1.2}
\hrulefill\\[-.6cm]
\begin{center}
$\begin{array}{rcllrclrrlll}
r(\nil) & = & \nil & \;  &
r(h.p) & = &  \nil + \nil  & \;  &  r(\mu.p) & = &  \mu .r(p) \mbox{ if $\mu \in L \cup \{\tau\}$}\\  
r(p + p') & = &   r(p) + r(p') & \quad &
r(p \para p') & = &   r(p) \para r(p') & \;  &
r(C) & = & C' \; \mbox{with $C' \eqdef r(p)$ if $C \eqdef p$}\\[-.2cm]
\end{array}$
$\begin{array}{rcllrcllllll}
\end{array}$
\end{center}}
\hrulefill
\caption{Restriction function} \protect\label{restr}
\end{table}
 
 \begin{table}[t]

{\renewcommand{\arraystretch}{1.2}
\hrulefill\\[-.7cm]
\begin{center}
$\begin{array}{rcllrclrrlll}
In(\nil) & = & \emptyset & \quad &
In(\mu.p) & = &  \{\mu\} & \quad & In(p + p') & = &   In(p) \cup In(p')\\
In(p \para p') & = &   In(p) \cup In(p') & \quad &
In(C) & = & In(p) \; \mbox{ if $C \eqdef p$}\\[-.2cm]
\end{array}$
$\begin{array}{rcllrcllllll}
\end{array}$
\end{center}}
\hrulefill
\caption{Function computing the initial actions} \protect\label{initial}
\end{table}

\begin{table}[t]
{\renewcommand{\arraystretch}{1.2}
\hrulefill\\[-.7cm]
\normalsize{

\begin{center}

$\begin{array}{cccccccc}
\bigfrac{}{(\nil, I):dni} &   &   \bigfrac{(p, I):dni, (q, I):dni, In(p+q) \subseteq L\cup \{\tau\}}{(p+q, I):dni}\\ 
\bigfrac{(p, I):dni, (q, I):dni}{(p \para q, I): dni}   &   & \bigfrac{(p, I):dni, p\neq\nil, (q, I):dni, E\vdash r(p)=r(q)}{(h.p + q, I):dni}\\
\end{array}$

$
\begin{array}{cccccccc}
 \bigfrac{(p, I):dni \quad \mu \in L \cup \{\tau\}}{(\mu.p, I):dni}  &   & \bigfrac{E\vdash p = \nil+\nil}{(h.p, I):dni}     &&
\bigfrac{\emptyset \neq In(p) \subseteq H, (p, I):dni}{(h.p, I):dni}\\  
\end{array}$

$
\begin{array}{cccccccc}
\bigfrac{C \in I}{(C, I): dni}  & & \bigfrac{C\not\in I, C \eqdef p, \{{\bf A_1-A_4}\} \vdash p= p', 
(p', I \cup \{C\}):dni}{(C, I): dni} \\
\end{array}$

\hrulefill\\
\end{center}}}
\caption{Typing proof system}\label{type}
\end{table}

Then we define a typing system on CFM processes such that, if a process $p$ is typed,
then $p$ satisfies DNI and, moreover, if $p$ is DNI, then there exists a process $p'$, obtained by possibly reordering
its summands via axioms ${\bf A_1- A_4}$, which is typed.

The typing system is outlined in Table \ref{type}, where
we are using a set $I$ of already scanned constants as an additional parameter (in order to avoid looping 
on recursively defined constants). A process $p$ is typed if $(p, \emptyset):dni$ is derivable
by the rules; this is often simply denoted by $p:dni$. The need for the argument $I$ is clear in the two 
rules for the constant $C$: if $C$ has been already scanned (i.e. $C \in I$), then $C$ is typed; otherwise
it is necessary to check that its body $p$ is typed (or possibly a sequential process $p'$ obtained by reordering its summands, i.e.,
such that $\{{\bf A_1-A_4}\} \vdash p =  p'$), using a set enriched by $C$ (condition $(p', I \cup \{C\}):dni$).

We are implicitly assuming
that the formation rules in this table respect the syntax of CFM; this means that, for instance, the second rule and the fourth one
require also that $p$ and $q$ are actually guarded processes because this is the case in $p + q$. Note that in the second rule
we are requiring that $In(p+q)$ is a subset of $L \cup \{\tau\}$, so that, as no high-level action is executable by $p$ and $q$ 
as their first action, no DNI check is required at this level of the syntax.

The interesting cases are the three rules about
the action prefixing operator and the rule about the choice operator when a summand starts with a high-level action. 
The fifth rule in Table \ref{type} states that if $p$ is typed, then also $\mu.p$ is typed, for each $\mu \in L\cup \{\tau\}$.
The sixth rule states that if $p$ is a deadlock place (condition $E \vdash p = \nil + \nil$), then $h.p$ is typed,
for each $h \in H$; note that $p$ cannot be $\nil$, because $h.\nil$ is not secure 
(the low-observable marking before performing $h$ has size one, while that after $h$
has size zero).
The seventh rule states that if $p$ is a typed term that can perform at least one action, but only high-level actions initially
(condition $\emptyset \neq In(p) \subseteq H$), then $h.p$ is typed. It is interesting to observe 
that $h.\tau.(\nil +\nil)$, which is DNI according
to Definition \ref{def-dni} based on branching team bisimilarity $\approx^\oplus$, is not DNI according to the current definition 
based on {\em rooted} branching team bisimilarity $\approx_c^\oplus$, and indeed $h.\tau.(\nil +\nil)$ cannot be typed.

The fourth rule in Table \ref{type} about the choice operator 
states that if we prefix a generic typed process $p$
with a high-level action $h$, then it is necessary that an additional typed summand $q$ is present such that 
$p \setminus H$ and $q \setminus H$ are rooted branching team equivalent; this semantic condition is expressed syntactically 
by requiring that $E \vdash r(p) = r(q)$, thanks to Theorem \ref{sound-complete-th}. Note that 
the condition $p \neq \nil$ is necessary, because $h.\nil + \nil$ is not secure; on the contrary, $h.\tau.\nil + \tau.\nil$ is typed 
because $E \vdash \tau.\nil = \tau.\nil$.
It is interesting to observe that this rule about the choice operator covers also the case when many summands start
with high-level actions. For instance, $h.l.l.\nil + (h.l.(h.l.\nil + l.\nil) + l.l.\nil)$ is typed because
$l.l.\nil$ and $h.l.(h.l.\nil + l.\nil) + l.l.\nil$ are typed and $E \vdash r(l.l.\nil) = r(h.l.(h.l.\nil + l.\nil) + l.l.\nil)$.
This strategy is intuitively correct (i.e., it respects DNI) because, by checking that the 
subterm $h.l.(h.l.\nil + l.\nil) + l.l.\nil$ is typed/DNI, we can safely ignore the other 
summand $h.l.l.\nil$, as it does not contribute any initial low-visible behavior.

Now we want to prove that the typing system characterizes DNI correctly.
To get convinced of this result, let us consider a couple of paradigmatic examples. 
The process constant  $C \eqdef h.l.C + l.C$ is DNI because, if $C \deriv{h} l.C$, then
$C \setminus H \approx_c (l.C)\setminus H$, which is equivalent to say that $r(C) \approx_c r(l.C)$.
Not surprisingly, we can type it. $(C, \emptyset):dni$ holds, because $(h.l.C + l.C, \{C\}):dni$ holds,
because, in turn, $(l.C, \{C\}):dni$ and $E \vdash r(l.C) = r(l.C)$.
On the contrary, $D \eqdef l.h.D$ is not DNI because if $h.D \deriv{h} D$, then $h.D \setminus H \not \approx_c D \setminus H$
as $h.D \setminus H$ is stuck, while $D\setminus H$ can perform $l$. As a matter of fact, $D$ is not typed:
to get $(D, \emptyset):dni$, we need $(l.h.D, \{D\}):dni$, which would require
$(h.D, \{D\}):dni$, which is false, as no rule for high-level prefixing is applicable.

\begin{proposition}
For each CFM process $p$, if $p:dni$, then $p$ satisfies DNI.
\proof By induction on the proof of $(p, \emptyset):dni$. It is enough to observe that for each rule, if we assume that the 
thesis holds on the premise conditions, then it also holds for the conclusion.
\fine
\end{proposition}

Note that the converse implication is not alway true, because of the ordering of summands. For instance,
$l.\nil + h.l.\nil$, which is clearly DNI, is not typed. However, $\{{\bf A_1-A_4}\} \vdash l.\nil + h.l.\nil =  h.l.\nil + l.\nil$,
where the process $h.l.\nil + l.\nil$ is typed.

\begin{proposition}
For each CFM process $p$, if $p$ is DNI, then there exists $p'$ such that
$\{{\bf A_1-A_4}\} \vdash p= p'$ and $p':dni$.
\proof 
The proof proceeds by induction on the structure of $p$, using a set $I$ of already scanned constants, in order to 
avoid looping on recursively defined constants. In other words, our induction is
on the net semantics for $p$, as defined in Table \ref{den-net-cfm}, where $I$ is initially empty and the base cases
are when $I = \const{p}$.

The base cases are $(\nil, I)$ and $(C, I)$ with $C \in I$. Both cases are trivial as these two terms generate nets that cannot do anything,
so that $(\nil, I)$ and $(C, I)$ are DNI and also typed, as required.

Case $(l.p, I)$: assume $(l.p, I)$
is DNI; then, also $(p, I)$ is DNI, and so, by induction $(p', I):dni$ for some $p'$ such that
$\{{\bf A_1-A_4}\} \vdash p = p'$. Then, $\{{\bf A_1-A_4}\} \vdash l.p = l.p'$ by substitutivity and moreover 
$(l.p', I):dni$ by the fifth rule in Table \ref{type}.

Case $(\tau.p, I)$: as above.

Case $(h.p, I)$: assume $(h.p, I)$
is DNI; then, this is possible only if either $p$ is a deadlock place or $(p, I)$ is DNI and $\emptyset \neq In(p) \subseteq H$.
In the former case, we can prove that $E \vdash p = \nil + \nil$ because the axiomatization is complete, so that we can use 
the sixth rule in Table \ref{type} to derive that $(h.p, I): dni$.
In the latter case, 
by induction, we know that $(p', I):dni$ for some $p'$ such that
$\{{\bf A_1-A_4}\} \vdash p = p'$. Then, $\{{\bf A_1-A_4}\} \vdash p = p'$
implies that $\emptyset \neq In(p') \subseteq H$.
Hence, we can use the seventh rule to derive $(h.p', I): dni$, where $\{{\bf A_1-A_4}\} \vdash h.p = h.p'$ follows by substitutivity.

Case $(C, I)$ with $C \not\in I$ and $C \eqdef p$:  assume $(C,I)$ is DNI. This means that also $(p, I \cup \{C\})$ is DNI.
By induction, we have that
$(p', I \cup \{C\}):dni$ for some $p'$ such that
$\{{\bf A_1-A_4}\} \vdash p = p'$. Now $(C, I):dni$
follows by the ninth rule in Table \ref{type}.

The only non-trivial inductive case is about summation. Assume that $(p_1 + p_2, I)$ is DNI. If $In(p_1 + p_2) \subseteq L \cup \{\tau\}$,
then both $(p_1, I)$ and $(p_2, I)$ are DNI; by induction, there exist $(p'_i, I):dni$  such that
$\{{\bf A_1-A_4}\} \vdash p_i= p_i'$ for $i = 1, 2$. Hence, $\{{\bf A_1-A_4}\} \vdash p_1+p_2= p_1'+p_2'$ by substitutivity; note that
also
$In(p_1' + p_2') \subseteq L \cup \{\tau\}$, so that 
also $(p'_1+p'_2, I):dni$ by the second rule in Table \ref{type}, as required.
On the contrary, if there exists $h \in In(p_1+p_2)$,
then there exist $q_1$ and $q_2$ such that 
 $\{{\bf A_1-A_4}\} \vdash p_1 + p_2 =  h.q_1 + q_2$. Since also $(h.q_1 + q_2, I)$ is DNI by (the adaptation of) Corollary \ref{cor-dni2} (to $\approx_c$),  
 it is necessary 
 that $q_1 \setminus H \approx_c q_2 \setminus H$, which is equivalent to $r(q_1) \approx_c r(q_2)$, 
in turn equivalent to stating that $E \vdash r(q_1) = r(q_2)$. 
Moreover, the DNI property has to be satisfied by $(q_1, I)$ and $(q_2, I)$. Hence, by induction,
 there exist $q_1', q_2'$ such that $\{{\bf A_1-A_4}\} \vdash q_i = q_i'$ and $(q_i', I):dni$, for $i = 1, 2$.
 By transitivity and substitutivity, we have  $\{{\bf A_1-A_4}\} \vdash p_1 + p_2 =  h.q'_1 + q'_2$. 
 Moreover, since $E \vdash r(q_1) = r(q_2)$, we also have that $E \vdash r(q'_1) = r(q'_2)$, 
 and so, by the proof system, $(h.q'_1 + q'_2, I):dni$, as required.
\fine
\end{proposition}

%
\section{Related Literature}\label{conc-bsec}
%

The non-interference problem in a distributed model of computation was first addressed in \cite{BG03,BG09}.
There, the Petri net class of {\em unlabeled elementary net systems} (i.e., safe, contact-free nets) was used to describe some  
information flow security properties, notably
{\em BNDC} (Bisimulation Non-Deducibility on Composition) and {\em SBNDC} (Strong BNDC), 
based on interleaving bisimilarity. These two properties do coincide on unlabeled elementary net systems,
but actually SBNDC is stronger on {\em labeled} FSMs; for instance, the CFM process
$l.h.l.\nil + l.\nil + l.l.\nil$ is BNDC \cite{FG01}, while it is not SBNDC; this explains why we have chosen 
SBNDC as our starting point towards the formulation of DNI.

In \cite{BG09} it is shown that BNDC can be characterized as a structural property of the elementary net concerning
two special classes of places: {\em causal places}, i.e., places for which there are an incoming high transition 
and an outgoing low transition; and {\em conflict places}, i.e., places for which there are both low-level and high-level outgoing transitions. 
The main theorem in \cite{BG09} states that if places of these types are not present or
cannot be reached from the initial marking, then the net is BNDC. 
Starting from \cite{BG09}, Baldan and Carraro in \cite{BC15} provide a 
{\em causal characterization} of BNDC on safe Petri nets (with injective labeling), in terms of the unfolding 
semantics. Nonetheless, the BNDC property is  based on an interleaving semantics
and the true-concurrency semantics is used only to provide efficient algorithms to check the possible presence of interferences.
For unbounded {\em partially observed} finite P/T nets (i.e., net with unobservable high transitions and {\em injective labeling} 
on low transitions), 
Best et al. proved in \cite{BDG10} that SBNDC is decidable.

Another paper studying non-interference over a distributed model is \cite{BHM15}. B\'erard et al. study 
a form of non-interference similar to SNNI \cite{FG01} for {\em High-level Message Sequence Charts} (HMSC), 
a scenario language for the description of distributed systems, based on composition of partial orders.
The model allows for unbounded parallelism and the observable semantics they use is interleaving and linear-time 
(i.e., language-based).
They prove that non-interference is undecidable in general, while it 
becomes decidable for regular behaviors, or for weaker variants based on observing local behavior only.
Also in this case, however, 
the truly-concurrent semantics based on partial orders is used mainly for algorithmic purpose; in fact, the authors 
shows that their decidable properties are PSPACE-complete, with procedures 
that never compute the interleaving semantics of the original HMSC.

 On the contrary, Baldan et al. in \cite{BBL17} define security policies, intuitively similar to non-interference, 
 where causality is used as a first-class concept. So, their notion of non-interference is
 more restrictive than those based on interleaving semantics. However, their approach is linear-time, while non-interference
 is usually defined on a branching-time semantics, i.e., on various forms of bisimulation. Moreover, it seems overly restrictive; for instance, the 
 CFM process $h.l.\nil + l.\nil$, which is DNI and so non-interferent according to our approach, would be considered insecure in their approach.
 
Finally, Joshua Guttman and Paul Rowe proposed \cite{GR15,Gut17} a model of distributed computation, called {\em frame}
(essentially a directed graph whose nodes are sequential systems, described by LTSs, and whose arcs are directed communication 
channels), onto which they define some 
form of intransitive non-interference, 
defined by exploiting the partial order description of the system executions. Hence, also their proposal 
advocates the use of truly-concurrent semantic models for defining non-interference. \\

\noindent
{\bf Acknowledgments:} 
This paper is a revised continuation of the conference paper \cite{Gor-petri20}, extending the approach to Petri nets with silent transitions.


\begin{thebibliography}{11}

\bibitem{ABS91}
C. Autant, Z. Belmesk, Ph. Schnoebelen,
\newblock  Strong bisimilarity on nets revisited, 
\newblock in Procs. PARLE'91, vol. II: Parallel Languages, LNCS 506, 295-312, Springer, 1991.


%
\bibitem{BC15}
P. Baldan, A. Carraro,
\newblock A causal view on non-intereference,
\newblock {\em Funda. Infor.} 140(1):1-38, 2015.

%
\bibitem{BBL17}
P. Baldan, A. Beggiato, A. Lluch-Lafuente,
\newblock Many-to-many information flow policies, 
\newblock 19th IFIP WG 6.1 {\em International Conference on Coordination Models and Languages}, 
LNCS 10319, 159-177, Springer, 2017.

%


%
\bibitem{BDG10}
E. Best, Ph. Darondeau, R. Gorrieri,
\newblock On the decidability of non-interference over unbounded Petri nets,
\newblock in Procs {\em 8th International Workshop on Security Issues in Concurrency} (SecCo 2010), EPTCS 51, 16-33, 2010.


%
\bibitem{BDKP91}
E. Best, R. Devillers, A. Kiehn, L. Pomello,
\newblock  Concurrent bisimulations in Petri nets,
\newblock {\em Acta Informatica}  28(3): 231-264, 1991.

%
\bibitem{BG03}
N. Busi, R. Gorrieri,
\newblock A survey on non-interference with Petri nets,
\newblock {\em Lectures on Concurrency and Petri Nets}, LNCS 3098, 328-344, Springer, 2003.

%
\bibitem{BG09}
N. Busi, R. Gorrieri,
\newblock Structural non-interference in elementary and trace nets,
\newblock {\em Mathematical Structures in Computer Science} 19(6): 1065-1090, 2009.

%
\bibitem{BHM15}
B. B\'erard, L. H\'elou\"et, J. Mullins,
\newblock Non-interference in partial order models,
\newblock in Procs ACSD'15, IEEE Computer Society,  80-89, 2015.

\bibitem{Ch93}
S. Christensen, 
\newblock {\em Decidability and Decomposition in Process Algebra},
\newblock Ph.D. Thesis, University of Edinburgh (1993)

%
\bibitem{DDM89}
P.~Degano, R.~De~Nicola, U.~Montanari,
\newblock Partial ordering descriptions and observations of nondeterministic concurrent systems,
\newblock in (J. W. de Bakker, W. P. de Roever, G. Rozenberg, Eds.)
\newblock {\em Linear Time, Branching Time and Partial Order in Logics and Models for Concurrency}, LNCS 354, 438-466, Springer, 1989.

%
\bibitem{Esp98}
J. Esparza,
\newblock Decidability and complexity of Petri net problems: An introduction,
\newblock  {\em Lectures on Petri Nets I: Basic Models},
\newblock   LNCS1491, 374-428, Springer, 1998.

\bibitem{FG95} R.~Focardi, R.~Gorrieri.
\newblock A classification of security properties.
\newblock {\it Journal of Computer Security} 3(1):5-33, 1995.


%
\bibitem{FG01}
R. Focardi, R. Gorrieri,
\newblock Classification of security properties (part I: Information flow),
\newblock in {\em Foundations of Security Analysis and Design I}, LNCS 2171, 331-396, Springer, 2001.

%



%
\bibitem{vGG89}
R.J. van Glabbeek, U. Goltz,
\newblock  Equivalence notions for concurrent systems and refinement of actions,
\newblock in Procs. MFCS'89, LNCS 379, 237-248, Springer, 1989.

\bibitem{vGW96}
R.J. van Glabbeek, W.P. Weijland,
\newblock  Branching time and abstraction in bisimulation semantics,
\newblock {\em Journal of the ACM} 43(3):555-600, 1996.

%
\bibitem{vG15}
R.J. van Glabbeek,
\newblock  Structure preserving bisimilarity: supporting an operational Petri net semantics of CCSP, 
\newblock in {\em Correct System Design} -- Symposium in Honor of Ernst-R\"udiger Olderog on the 
Occasion of His 60th Birthday, LNCS 9360, 99-130, Springer, 2015.

\bibitem{GM} J.A.~Goguen, J.~Meseguer.
\newblock Security policy and security models.
\newblock Proc.\ of Symposium on Security and Privacy (SSP'82), IEEE CS Press, pp.~11-20, 1982.


%

%
\bibitem{GV15}
R. Gorrieri, C. Versari,
{\em Introduction to Concurrency Theory: Transition Systems and CCS},
EATCS Texts in Theoretical Computer Science, Springer, 2015.

%
\bibitem{Gor17}
R.~Gorrieri,
\newblock {\em Process Algebras for Petri Nets: The Alphabetization of Distributed Systems},  
\newblock EATCS Monographs in Theoretical Computer Science, Springer, 2017.

\bibitem{Gor17b}
R.~Gorrieri,
\newblock Team bisimilarity, and its associated modal logic, for BPP nets, 
\newblock {\em Acta Informatica}, 2020. DOI: 10.1007/s00236-020-00377-4

\bibitem{Gor19b}
R.~Gorrieri,
\newblock Team equivalences for finite-state machines with silent moves, 
\newblock {\em Information and Computation} 275:104603, 2020. DOI:10.1016/j.ic.2020.104603

%

 \bibitem{Gor-petri20}
 R. Gorrieri,
\newblock Interleaving vs true concurrency: some instructive security examples, 
\newblock in Procs. Petri Nets 2020, LNCS 12152, 131-152, Springer, 2020.


\bibitem{Gor22-tcs}
R.~Gorrieri,
\newblock A study on team bisimulation and h-team bisimulation for BPP nets, 
{\em Theoretical Computer Science} 897:83-113, 2022.

\bibitem{Gor21}
R.~Gorrieri,
\newblock Place bisimilarity is decidable, indeed!,
\newblock arXiv:2104.01392, april 2021.

\bibitem{Gor21b}
R. Gorrieri,
\newblock Branching place bisimilarity: A decidable behavioral equivalence for finite Petri nets with silent moves,
\newblock in Procs. {\em 41$^{st}$ Formal Techniques for Distributed Objects, Components, and Systems} (FORTE'21), 
LNCS 12719, 80-99, Springer, 2021.


\bibitem{Gor23b}
R.~Gorrieri,
\newblock Branching place bisimilarity, 
\newblock  CoRR abs/2305.04222, September 2023, {https://arXiv.2305.04222}

\bibitem{GR15}
Joshua D. Guttman, Paul D. Rowe,
\newblock A cut principle for information flow,
\newblock in Procs. IEEE Computer Security Foundations (CSF 2015), 107-121, IEEE Computer Society Press, 2015.

\bibitem{Gut17}
Joshua D. Guttman,
\newblock Information flow, distributed systems, and refinement, by example,
\newblock {\em Concurrency, Security, and Puzzles: Essays Dedicated to Andrew William Roscoe on the 
Occasion of His 60th Birthday}, 
LNCS 10160, 88-103, Springer, 2017.

%
\bibitem{Hoa85}
C.A.R. Hoare,
\newblock {\em Communicating Sequential Processes},
\newblock Prentice-Hall International Series in Computer Science, 1985.


%
\bibitem{HMU01}
J.E. Hopcroft, R. Motwani, J.D. Ullman,
\newblock {\em Introduction to Automata Theory, Languages and Computation}, 2nd ed.,
\newblock Addison-Wesley, 2001.

%
\bibitem{JGKW20}
D.N. Jansen, J.F. Groote, J.J.A. Keiren, A. Wijs. 
An $O(m \cdot log\, n)$ algorithm for branching bisimilarity on labelled transition systems,
in Procs. TACAS'20, LNCS 12079, 3-20, Springer, 2020.

%


%



%
\bibitem{Mil89} R. Milner. {\it Communication and Concurrency},
Prentice-Hall, 1989.


%
\bibitem{NT84}
M. Nielsen, P.S. Thiagarajan,
\newblock  Degrees of non-determinism and concurrency: A Petri net view,
\newblock in Procs. of the Fourth Conference on Foundations of Software Technology and Theoretical Computer Science (FSTTCS'84),
LNCS 181, 89-117, Springer, 1984.

%

%


%


%
\bibitem{Pet81}
J.L. Peterson,
\newblock {\em Petri Net Theory and the Modeling of Systems}, Prentice-Hall, 1981.

\bibitem{Rei85}
W. Reisig,
\newblock {\em Petri Nets: An Introduction},
\newblock EATCS Monographs in Theor. Comp. Science,
\newblock Springer, 1985.


%
\bibitem{Ry01}
P. Ryan,
\newblock Mathematical models of computer security,
\newblock in {\em Foundations of Security Analysis and Design I}, LNCS 2171, 1-61, Springer, 2001.

\bibitem{RS} P.~Ryan, S.~Schneider.
Process algebra and noninterference, in Proc.\ of 12th Computer
Security Foundations Workshop (CSFW),  214-227, IEEE CS Press, 1999.

\bibitem{RT88}
A. Rabinovich,  B.A. Trakhtenbrot,
\newblock Behavior structures and nets,
\newblock {\em Fundamenta Informaticae} 11(4):357-404, 1988.


\end{thebibliography}
\end{document}